\documentclass[10pt,letterpaper]{article}
\usepackage[top=0.85in,left=2.75in,footskip=0.75in]{geometry}

\usepackage{amsmath,amssymb}

\usepackage{changepage}

\usepackage[utf8x]{inputenc}

\usepackage{textcomp,marvosym}

\usepackage{cite}

\usepackage{nameref,hyperref}

\usepackage[right]{lineno}

\usepackage{microtype}
\DisableLigatures[f]{encoding = *, family = * }

\usepackage[table]{xcolor}

\usepackage{array}

\newcolumntype{+}{!{\vrule width 2pt}}

\newlength\savedwidth

\raggedright
\usepackage[aboveskip=1pt,labelfont=bf,labelsep=period,justification=raggedright,singlelinecheck=off]{caption}

\makeatletter
\renewcommand{\@biblabel}[1]{\quad#1.}
\makeatother

\usepackage{lastpage,fancyhdr,graphicx}
\usepackage{epstopdf}
\fancyheadoffset[L]{2.25in}
\fancyfootoffset[L]{2.25in}
\usepackage{xspace}
\newcommand*{\eg}{e.g.\@\xspace}

\newcommand*{\ie}{i.e.\@\xspace}
\newcommand*{\resp}{resp.\@\xspace}
\newcommand*{\vs}{vs.\@\xspace}

\newcommand*{\pastmeridiem}{p.m.\@\xspace}
\newcommand\Fscore{$F_1$-score}
\reversemarginpar

\usepackage{mathptmx}
\usepackage{amsmath}
\usepackage{amssymb}
\usepackage{amsthm}
\usepackage{siunitx}
\usepackage{url}
\usepackage{svg}
\begin{document}
\vspace*{0.2in}

\title{Robust detection in bioacoustic sensor networks}

\begin{flushleft}
{\Large
\textbf\newline{Robust sound event detection in bioacoustic sensor networks} 
}
\newline
\\
Vincent Lostanlen\textsuperscript{1,2,3*},
Justin Salamon\textsuperscript{2,3},
Andrew Farnsworth\textsuperscript{1},
Steve Kelling\textsuperscript{1}, and
Juan Pablo Bello\textsuperscript{2,3}
\\
\bigskip
\textbf{1} Cornell Lab of Ornithology, Cornell University, Ithaca, NY, USA
\\
\textbf{2} Music and Audio Research Laboratory, New York University, New York, NY, USA
\\
\textbf{3} Center for Urban Science and Progress, New York University, New York, NY, USA
\\
\bigskip

%
%





* vincent.lostanlen@nyu.edu

\end{flushleft}

\section*{Abstract}
Bioacoustic sensors, sometimes known as autonomous recording units (ARUs), can record sounds of wildlife over long periods of time in scalable and minimally invasive ways.
Deriving per-species abundance estimates from these sensors requires detection, classification, and quantification of animal vocalizations as individual acoustic events.
Yet, variability in ambient noise, both over time and across sensors, hinders the reliability of current automated systems for sound event detection (SED), such as convolutional neural networks (CNN) in the time-frequency domain.
In this article, we develop, benchmark, and combine several machine listening techniques to improve the generalizability of SED models across heterogeneous acoustic environments.
As a case study, we consider the problem of detecting avian flight calls from a ten-hour recording of nocturnal bird migration, recorded by a network of six ARUs in the presence of heterogeneous background noise.
Starting from a CNN yielding state-of-the-art accuracy on this task, we introduce two noise adaptation techniques, respectively integrating short-term (\SI{60}{\milli\second}) and long-term (\SI{30}{\minute}) context.
First, we apply per-channel energy normalization (PCEN) in the time-frequency domain, which applies short-term automatic gain control to every subband in the mel-frequency spectrogram.
Secondly, we replace the last dense layer in the network by a context-adaptive neural network (CA-NN) layer, \ie{} an affine layer whose weights are dynamically adapted at prediction time by an auxiliary network taking long-term summary statistics of spectrotemporal features as input.
We show that PCEN reduces temporal overfitting across dawn \vs{} dusk audio clips whereas context adaptation on PCEN-based summary statistics reduces spatial overfitting across sensor locations.
Moreover, combining them yields state-of-the-art results that are unmatched by artificial data augmentation alone.
We release a pre-trained version of our best performing system under the name of BirdVoxDetect, a ready-to-use detector of avian flight calls in field recordings.

\section*{Introduction}

\subsection*{Machine listening for large-scale bioacoustic monitoring}

The past decades have witnessed a steady decrease in the hardware costs of sound acquisition \cite{segura2015ieeesensors}, processing \cite{mack2015spectrum}, transmission \cite{hecht2016spectrum}, and storage \cite{mccallum2017web}.
As a result, the application domain of digital audio technologies has extended far beyond the scope of interhuman communication to encompass the development of new cyberphysical systems \cite{stowell2015detection}.
In particular, passive acoustic sensor networks, either terrestrial or underwater, contribute to meet certain challenges of industrialized societies, including wildlife conservation \cite{laiolo2010biolconserv}, urban planning \cite{bello2018chapter}, and the risk assessment of meteorological disasters \cite{zhao2014jphysoceanogr}.

Biodiversity monitoring is one of the most fruitful applications of passive acoustics.
Indeed, in comparison with optical sensors, acoustic sensors are minimally invasive \cite{merchant2015methecolevol}, have a longer detection range
--- from decameters for a flock of migratory birds to thousands of kilometers for an oil exploration airgun \cite{nieukirk2012jasa} --- and their reliability is independent of the amount of daylight \cite{blumstein2011applecol}.
In this context, one emerging application is the species-specific inventory of vocalizing animals \cite{marques2013biolrev}, such as birds \cite{shonfield2017ace}, primates \cite{heinicke2015mee}, and marine mammals \cite{baumgartner2013jasa}, whose occurrence in time and space reflects the magnitude of population movements \cite{stewart2018ecosphere}, and can be correlated with other environmental variables, such as local weather \cite{oliver2018science}.

The principal motivation for this article is to monitor bird migration by means of a bioacoustic sensor network \cite{fiedler2009ringmigr}.
From one year to the next, each species is susceptible to alter its migratory onset and route as a function of both intrinsic factors \cite{gordo2007bird} and extrinsic (e.g. human-caused) environmental pressures \cite{bairlein2016science,loss2015annrev}.
Mapping in real time \cite{dokter2018nature}, and even forecasting \cite{farnsworth2014ai,vandoren2018science}, the presence and quantity of birds near hazardous sites (\eg{} airports \cite{devault2011interspecific}, windfarms \cite{drewitt2006ibis}, and dense urban areas \cite{blair1996ecolappl}) would enable appropriate preventive measures for avian wildlife conservation, such as temporary reduction of light pollution \cite{vandoren2017pnas}.
In addition, it could benefit civil aviation safety as well as agricultural planning \cite{bauer2017bioscience}.

At present, monitoring nocturnal bird migration is a challenge of integrating complementary methods to try to produce the most comprehensive understanding of migrants' movements.
The two most readily available sources of information for tracking the movements of avian populations at large (e.g. continental) scales are weather surveillance radar data \cite{farnsworth2016ecolappl} and crowdsourced observations of birders \cite{sullivan2014biolconserv}.
Both of these information sources are valuable but imperfect.
In particular, the former does not distinguish different species, rather providing data only on bird biomass aloft.
Conversely, the latter is dominated by diurnal information, which does not describe spatial and temporal distribution of species when they are actively migrating at night, and is sparse, requiring state-of-the-art computational approaches to produce distribution models.
In contrast, flight calls can provide species information, at the least for vocal species; and may, in principle, be detected in real time \cite{farnsworth2005auk}.
Supplementing spatiotemporal exploratory models \cite{fink2010spatiotemporal}, currently restricted to radar and observational modalities \cite{fink2014ai}, with the output of a bioacoustic sensor network, could improve our ability to detect species flying over the same area simultaneously, and offer new insights in behavioral ecology and conservation science.

In a large-scale setting of bioacoustic monitoring at the continental scale and over multi-month migration seasons, the task of counting individual vocalizations in continuous recordings by human annotators to achieve these ends is impractical, unsustainable, and unscalable.
Rather, there is dire need for a fully automated solution to avian flight call detection \cite{pamula2017osa}, that would rely on machine listening, \ie{} the auditory analogue of computer vision \cite{stowell2018computational}.
We propose that, in the future, each sensor could run autonomously \cite{ross2018ecologicalresearch}, by sending hourly digests of bird vocalization activity to the central server, which in turn would aggregate information from all sensors, ultimately resulting in a spatiotemporal forecast of nocturnal migration \cite{shamoun2016plosone}.

The detection of far-field signals despite the presence of background noise constitutes a fundamental challenge for bioacoustic sensor networks \cite{warren2006animalbehaviour}.
Whereas, in a typical fieldwork setting, a human recordist would use a directional microphone and point it towards a source of interest, thus minimizing background noise or other interference \cite{lanzone2009revealing}, autonomous recording units (ARUs) are most often equipped with single omnidirectional microphones \cite{hobson2002wildlife}.
As a result, instead of tracking the sources of interest, they capture a global \emph{soundscape} (sonic landscape) of their environment \cite{pijanowski2011bioscience}, which also comprises spurious sources of noise \cite{naguib2003jasa}.
Furthermore, in the context of avian flight calls, migratory birds move rapidly, vocalize intermittently, and may simultaneously be present at multiple azimuths around a sensor \cite{krim1996spm}.
Consequently, none of the well-established methods for beamforming animal vocalizations --- which assume that each sensor combines multiple directional microphones \cite{wilson2018ace} --- would apply to the use case of flight call monitoring.
On the contrary, we formulate a scenario in which sound event detection occurs in natural soundscapes without prior localization of sources.
This formulation represents a potential use case for the deployment of a large-scale bioacoustic sensor network consisting of low-cost, single-microphone hardware \cite{mydlarz2017appliedacoustics}.

Because migratory birds appear to vocalize at a low acoustic intensity and at a high distance to the sensor \cite{knight2018bioacoustics}, simple energy-based detection functions \cite{evans2005passenger} or spectrotemporal template matching \cite{kaewtip2016jasa} may be inadequate for solving problems of retrieving avian flight calls in continuous recordings.
Instead, machine learning appears necessary for detecting acoustic events in noisy, highly reverberant environments \cite{heittola2018chapter}.
Yet, one fundamental assumption behind conventional machine learning methods is that samples from the training set and samples from the test set are drawn from the same high-dimensional probability distribution.

In the specific case of bioacoustic sensor networks, a training set may consist of audio clips from a limited number of recordings that are manually annotated a priori, whereas the test set will encompass a broader variety of recording conditions, including days, sensor locations, and seasons that are unreviewed or unlabeled \cite{joly2017clef}.
Although it is plausible to assume that, from one recording condition to another, the statistical properties of the flight calls themselves --- hereafter denoted as foreground --- are identically or almost identically distributed, the same cannot be said of background sources of noise.
Rather, natural soundscapes, even at the spatial scale of a few square kilometers and at the temporal scale of a few hours, may exhibit large variations in background noise spectra \cite{ulloa2018ecologicalindicators}.
Therefore, state-of-the-art machine learning systems for sound event detection, once trained on the far-field recordings originating from a limited number of sensors, might fail to generalize once deployed on a different sensor \cite{brumm2017mee}.

The crux of the challenge of robust sound event detection resides in the practical limitations of human annotation.
In a supervised setting, the diversity of recording conditions that are available for training the sound event detection system at hand is necessarily lower than those on which the same system will eventually be deployed.
The current lack of robust methods for sound event detection in heterogeneous environments have caused past bioacoustic studies to focus on relatively few acoustic sensors in close proximity \cite{marcarini2008icassp,efford2009ecology}.
Nevertheless, the goal of deploying a large-scale network of acoustic sensors for avian migration monitoring requires sound event detection to adapt to nonstationarities (\ie{} variations in time) and nonuniformities (\ie{} variations in space) of background noise.
In this article, we propose a combination of novel methods, not only to improve the accuracy of state-of-the-art detectors on average, but also to make these detectors more reliable across recording conditions, such as those arising at dawn \vs{} dusk or across different sensor locations.

\subsection*{Evidence of technical bias in state-of-the-art bioacoustic detection systems}

\begin{figure}
\begin{minipage}{1.0\linewidth}
\centering
\centerline{\includegraphics[width=0.75\linewidth]{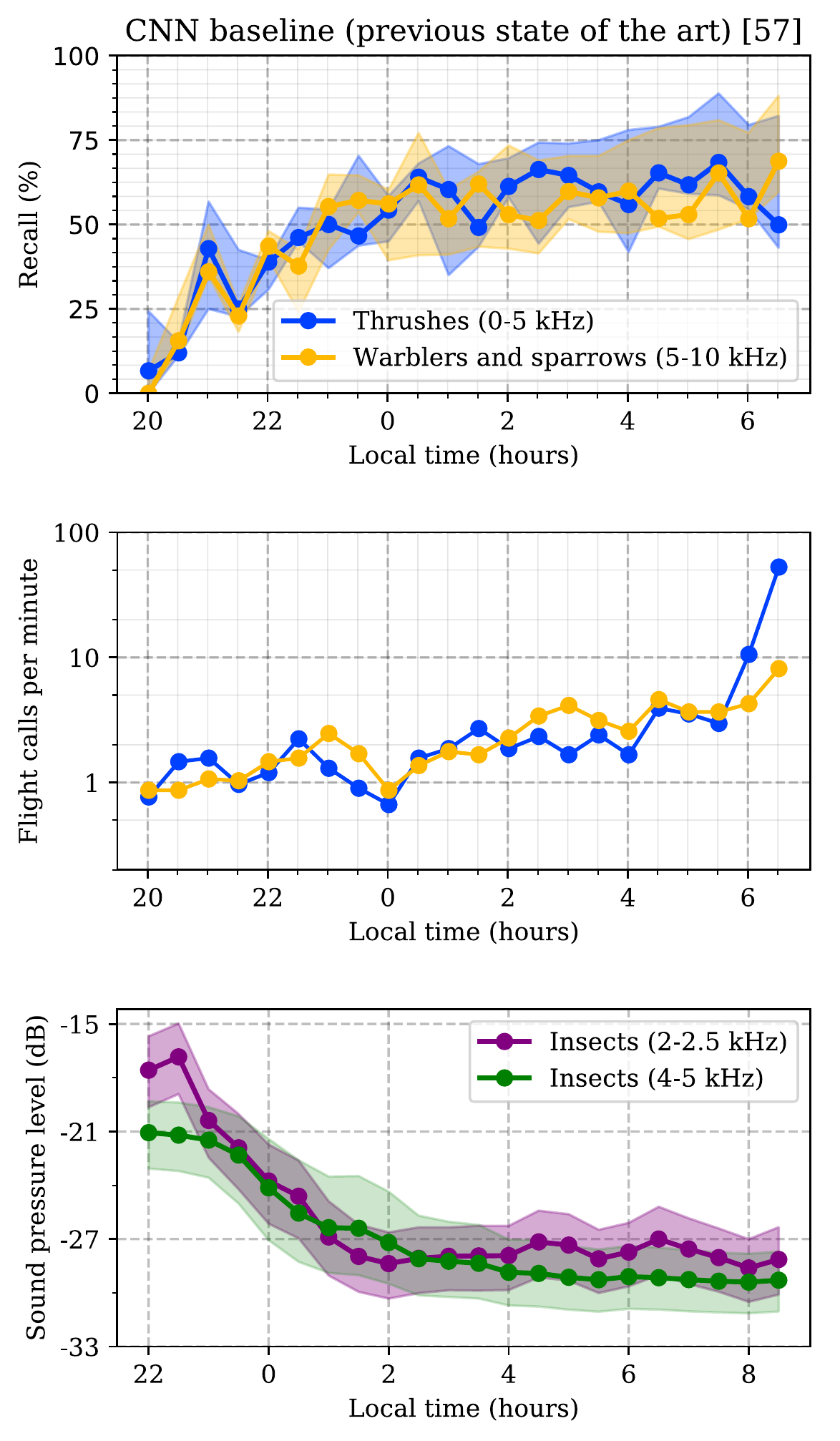}}
\end{minipage}
\caption{On BirdVox-full-night, the recall of the baseline CNN increases with time, as the density of flight calls increases and the noise level decreases. Shaded areas denote interquartile variations across sensors.}
\label{fig:icassp-sota}
\end{figure}

For example, Fig \ref{fig:icassp-sota} illustrates the challenges of a state-of-the-art sound event detector of nocturnal flight calls, namely the convolutional neural network architecture of \cite{salamon2017icassp}, hereafter called ``CNN baseline'' in this paper.
In the top plot, which is replicated from a previous study \cite{lostanlen2017icassp}, the authors measured the evolution of recall of the CNN baseline over a publicly available machine listening benchmark for avian flight call detection, named BirdVox-full-night.
This benchmark consists of six continuous recordings of flight calls, corresponding to six different autonomous recording units; it will be described in further detail in the Methods section.
Over the course of ten hours, the CNN baseline system exhibits large variabilities in recall, \ie{} fraction of detected events that are true positives, through time: both within the vocal ranges of thrushes (from $0$ to $\SI{5}{\kilo\hertz}$) and of warblers and sparrows (from $5$ to $\SI{10}{\kilo\hertz}$), recall oscillates between $5\%$ and $35\%$ during dusk and night before soaring rapidly up to $75\%$.

One explanation for these variations lies in the unequal amount of available training data in function of recording conditions: as shown in the middle plot of Fig \ref{fig:icassp-sota}, the average number of flight calls per minute increases with time.
Because its loss function assigns the same importance to every misclassified example, the baseline CNN model overfits dawn audio clips and underfits dusk audio clips.

The nonstationarity of background noise, at the time scale of a full night, aggravates the phenomenon of overfitting of this machine learning system.
In the bottom plot of Fig \ref{fig:icassp-sota}, we extract the evolution of sound pressure level (SPL) within a narrow subband of interest (between $2$ and $\SI{2.5}{\kilo\hertz}$), as well as within the subband corresponding to its second harmonic (\ie{} between $4$ and $\SI{5}{\kilo\hertz}$).
Both correspond to the frequency range of stridulating insects in the background.
In both subbands, we find that the median SPL, as estimated over $30$-minute temporal windows, decreases by about $\SI{10}{\decibel}$ between $8$ \pastmeridiem and midnight.
This is because insect stridulations are most active at dusk, before fading out gradually.

The large variations in accuracy through time exhibited above are particularly detrimental when applying the baseline CNN detector for bird migration monitoring.
Indeed, deploying this baseline CNN detector over a bioacoustic sensor network will likely lead to a systematic underestimation of vocal activity of migratory birds at dusk and an overestimation at dawn.
This is a form of technical bias that, if left unchecked, might lead to wrong conclusions about the behavioral ecology and species composition of nocturnally migrating birds.
Furthermore, and perhaps worse, such bias could create a foundation for conservation science that, contrary to original ambitions, is not based on the actual distribution and attributes of the target species of concern.

\subsection*{Contributions}
The aim of this article is to improve the reliability of state-of-the-art sound event detection algorithms across acoustic environments, thus mitigating the technical bias incurred by nonstationarity and nonuniformity in background noise.
We present four contributions to address this problem.

First, we develop a new family of neural network architectures for sound event detection in heterogeneous environments.
The commonality among these architectures is that they comprise an auxiliary subnetwork that extracts a low-dimensional representation of background noise and incorporates it into the decision function of the main subnetwork.
As such, they resemble context-adaptive neural networks (CA-NNs), \ie{} an existing line of research in automatic speech recognition from multichannel audio input \cite{delcroix2015icassp}.
Yet, our CA-NN architectures differ from the current literature, both in the choice of auxiliary features and in the choice of mathematical formulation of the context-adaptive layer.
We introduce long-term spectral summary statistics as auxiliary features for representing acoustic environments, whereas previous publications \cite{huemmer2017icassp} relied on short-term spatial diffuseness features \cite{schwarz2015icassp}.
Furthermore, we generalize the mathematical formulation of context adaptation --- initially described as a mixture-of-experts multiplicative gate \cite{delcroix2018context} --- within the broader topic of dynamic filter networks \cite{jia2016nips}, and especially discuss the cases of context-adaptive dense layers with dynamic weights or with dynamic biases.

Second, we apply a new time-frequency representation to bioacoustic signal detection.
Known as per-channel energy normalization (PCEN), this representation was recently proposed with the aim of improving robustness to channel distortion in a task of keyword spotting \cite{wang2017icassp}.
In this article, we demonstrate that, after we reconfigure its intrinsic parameters appropriately, PCEN also proves to be useful in outdoor acoustic environments.
Indeed, we find that it enhances transient events (\eg{} bird vocalizations) while discarding stationary noise (\eg{} insect stridulations).
To the best of our knowledge, this article is the first in successfully applying PCEN to the analysis of non-speech data.

Third, we conduct a thorough evaluation of the respective effects of each component in the development of a deep convolutional network for robust sound event detection: presence of artificial data augmentation; choice of time-frequency representation (PCEN \vs{} logarithm of the mel-frequency spectrogram); and formulation of context adaptation.
The overall computational budget that is incurred by this thorough evaluation is of the order of $10$ CPU-years.
After summarizing the results of our benchmark, we provide conclusive evidence to support the claim that CA-CNN and PCEN, far from interchangeable, are in fact complementary.
Our experiments demonstrate that context adaptation alone fails to improve the generalizability of a logmelspec-based deep learning model for avian flight call detection. 
A positive result contrasts this negative result: after replacing logmelspec by PCEN, context adaptation improves the generalizability of the detector, and this improvement is unmatched by data augmentation alone.

Finally, we combine all our findings into a deep learning system for avian flight call detection in continuous recordings.
This system is named BirdVoxDetect, is written in the Python language, and is released under the MIT free software license.
This open source initiative is directed towards the machine listening community, in order to allow the extension of our research beyond its current application setting.
In addition, we release our best performing BirdVoxDetect model under the form of a command-line interface, which segments and exports all detected sound events as separate audio clips, thus facilitating further inspection or automatic processing.
This interface is directed towards the avian bioacoustics community, in order to allow the large-scale deployment of autonomous recording units for flight call monitoring.
In an effort of conducting transparent, sustainable, and reproducible audio research \cite{mcfee2019spm}, BirdVoxDetect also comprises documentation, a test suite, a Python package indexation, and an interoperable application programming interface (API).
The source code of BirdVoxDetect is freely available at the following URL address: \url{https://github.com/BirdVox/birdvoxdetect}.

The development stage of an automated solution for bioacoustic monitoring typically relies on a small, prototypical subset of the sensor network.
In this article, so as to estimate the ability of the system to generalize beyond this prototypical subset, we propose a ``leave-one-sensor-out'' cross-validation methodology. Furthermore, because all sensors in the BirdVox-full-night dataset consist of identical acquisition hardware, the experimental benefit of integrating long-term spectrotemporal summary statistics as auxiliary features to our context-adaptive neural network can be interpreted as a form of robustness to spatial nonuniformity of environmental noise, and not merely as robustness to incidental variations in the impedance curve of each sensor. 

No previous publication has investigated the relational effect of PCEN and CA, so this result is novel and unanticipated. 
Although the number of papers which propose to employ PCEN for acoustic event detection has grown in recent months, all of them have motivated PCEN by appealing to a property of robustness to loudness variations, not robustness to artificial data augmentation.
The fact that deep convolutional networks for sound event detection benefit more from data augmentation in the time-frequency domain if the logmelspec frontend is replaced by PCEN is a novel and unanticipated finding of our article.

\subsection*{Related work}

To the best of our knowledge, the only computational system for long-term bird migration monitoring that currently relies on acoustic sensor data is Vesper \cite{mills2018zenodo}.
In order to detect thrushes, warblers, and sparrows, Vesper implements algorithms originally described in \cite{evans2005passenger} that do not adapt dynamically to the changes in background noise described above.
Instead, these detectors employ a measure of spectral flux \cite{klapuri1999icassp} within manually defined passbands, associated with some \emph{ad hoc} constraints on the minimal and maximal duration of a flight call.
This straightforward and computationally elegant approach has been a standard for many in the amateur, academic, and professional migration monitoring communities. Yet, despite its simplicity and computational efficiency, such algorithms suffer from considerable shortcomings in detection accuracy, and may not be a reliable replacement for human inspection.
In particular, a previous evaluation campaign showed that these detectors can exhibit precision and recall metrics both below $10\%$ in a multi-sensor setting \cite{lostanlen2017icassp}.

Another line of research that is related to this article is that of ``bird detection in audio'' \cite{stowell2016mlsp}, \ie{} a yearly challenge during which machine listening researchers train systems for general-purpose detection of vocalizations over a public development dataset, and then compete for maximal accuracy over a private evaluation dataset.
In recent years, the organizers of this challenge have managed to attract researchers from the machine learning and music information retrieval (MIR) communities \cite{stowell2018mee}.
This had led to the publication of new applications of existing machine learning methods to the domain of avian bioacoustics: these include multiple instance learning \cite{grill2017eusipco}, convolutional recurrent neural networks \cite{cakir2017eusipco}, and densely connected convolutional networks \cite{pellegrini2017eusipco}.
Despite its undeniable merit of having gathered several data collection initiatives into a single cross-collection evaluation campaign, the methodology of the ``bird detection in audio'' challenge suffers from a lack of interpretability in the discussion of results \emph{post hoc}.
Indeed, because bird vocalizations are not annotated at the time scale of individual acoustic events but at the time scale of acoustic scenes, it is impossible to draw a relationship of proportionality between the average miss rates of competing systems and their respective technical biases, in terms of robustness to nonstationarity and nonuniformity of background noise.
Furthermore, because these acoustic scenes are presented to the competitors under the forms of ten-second audio segments, rather than continuous recordings of several hours, the development and evaluation of some context-adaptive machine listening methods, such as the ones relying on spectrotemporal summary statistics for modeling background noise, remain out of the scope of practical applicability.

Until very recently, the scope of applications of per-channel energy normalization (PCEN) in deep learning was restricted to speech and music applications \cite{schluter2018ismir,millet2019icassp}.
This situation changed over the past six months with the publication of \cite{lostanlen2018spl} which conducts a statistical analysis of PCEN in various outdoor environments; and \cite{zinemanas2019fruct}, which trains a PCEN layer within an end-to-end urban sound event detection system.
Furthermore, one publication has raised the idea of using PCEN in a deep learning system for a task of bird species identification \cite{kahl2018clef}, but did not report any result.
Lastly, one publication reports unsuccessful results from using PCEN in the same task of bird species identification \cite{schluter2018clef}.
This article is the first in reporting significant improvements from replacing logmelspec by PCEN in a task of bioacoustic sound event detection; the first in demonstrating the importance of using PCEN in context-adaptive machine listening; and the first in showing experimentally that PCEN improves robustness to nonstationarity and nonuniformity of background noise at the scale of an entire acoustic sensor network over time scales of multiple hours.

\section*{Methods}

\subsection*{Overview}

\begin{figure}
\begin{minipage}{1.0\linewidth}
\centering
\centerline{\includegraphics[width=1.0\linewidth]{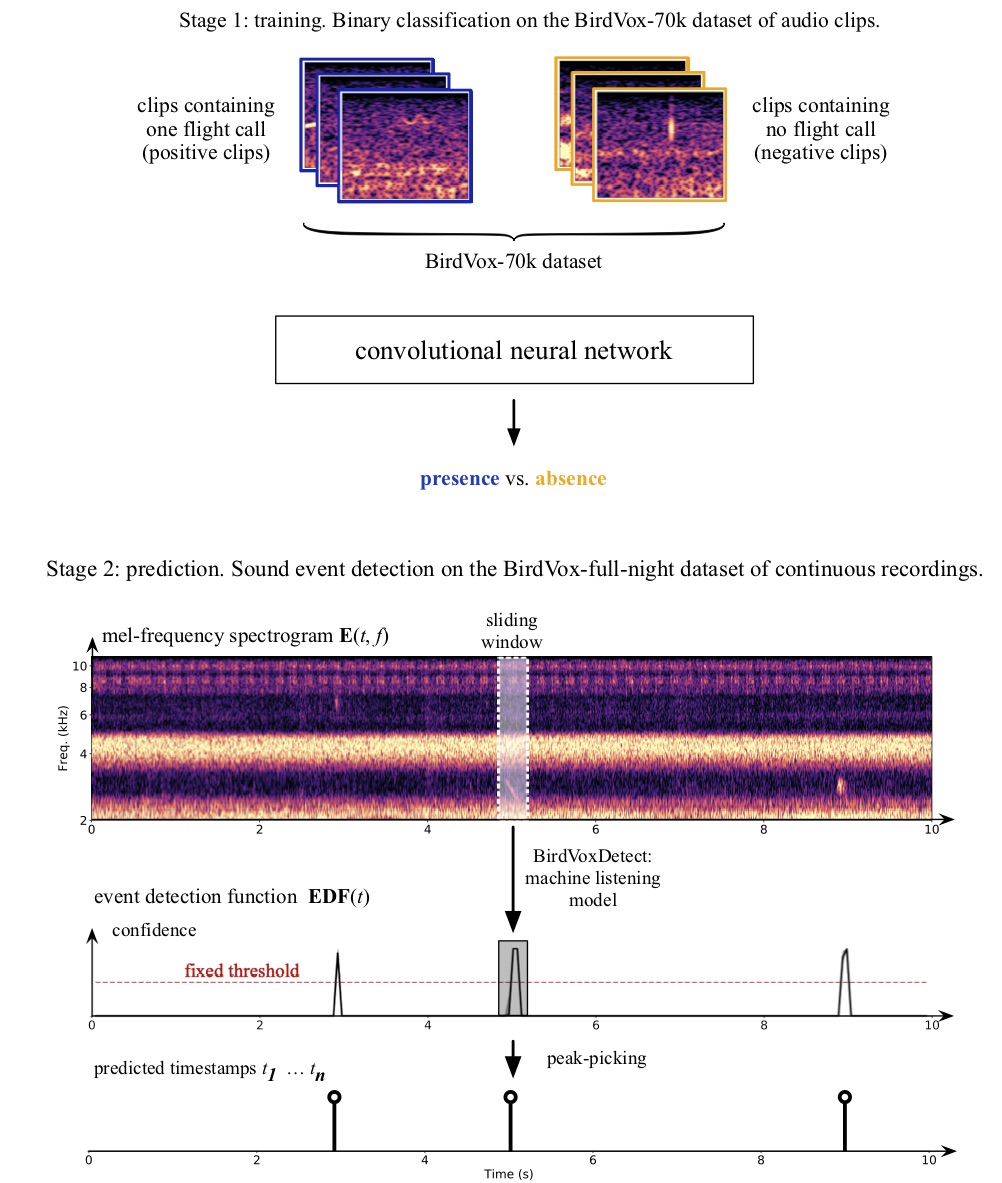}}
\end{minipage}
\caption{Overview of the presented baseline. After training a deep learning model to identify the presence of a sound event within short audio clips (\SI{150}{\milli\second}), we run this model on a continuous recording by a sliding window procedure. We compare the peaks in the resulting event detection function (EDF) with a fixed threshold $\tau$ in order to obtain a list of predicted timestamps for the sound event of interest. In the case of the presented baseline, these sound events of interest are avian flight calls; the input representation is a mel-frequency spectrogram; and the deep learning model is a convolutional network.}
\label{fig:binary-classification-diagram}
\end{figure}

All methods presented herein rely on machine learning.
Therefore, their comparison entails a training stage followed by a prediction stage.
Fig \ref{fig:binary-classification-diagram} illustrates both stages schematically.

First, we formulate the training stage as binary classification of presence \vs{} absence of a sound event.
In this setting, the input to the system is a short audio clip, whose duration is equal to $\SI{150}{\milli\second}$.
We represent this audio clip by a time-frequency representation $\mathbf{E}(t,f)$.
In the state-of-the-art model of \cite{lostanlen2017icassp}, the matrix $\mathbf{E}(t,f)$  contains the magnitudes in the mel-frequency spectrogram near time $t$ and mel frequency $f$.
The output to the system is a number $y$ between $0$ and $1$, denoting the probability of presence of a sound event of interest.
In the case of this paper, this sound event is a nocturnal flight call.

Next, we formulate the prediction stage as sound event detection.
In this setting, the input to the system is an acoustic scene of arbitrarily long duration.
The output of the system is an event detection function $y(t)$, sampled at a rate of $20$ frames per second.
For every $t$, we compute $y(t)$ by sliding a window of duration equal to $\SI{150}{\milli\second}$ and hop size equal to $\SI{50}{\milli\second}$ over the time-frequency representation $\mathbf{E}(t,f)$ of the acoustic scene.
We turn the event detection function $y(t)$ into a list of predicted timestamps by a procedure of thresholding and peak extraction.
The total number of predicted timestamps is a computer-generated estimate of the vocal activity of migratory birds near the sensor location at hand.
In the realm of avian ecology, this number could potentially be used as a proxy for the density of birds over the course of an entire migration season.
Furthermore, the short audio clips corresponding to detected flight calls could be subsequently passed to an automatic species classifier \cite{salamon2017icassp} to obtain the distribution of species in the vicinity of each sensor.

We shall describe the procedures of training and evaluating our proposed system in greater detail in the Experimental Design section of this article.

\subsection*{Context-adaptive neural network}

\subsubsection*{Related work}
There is a growing body of literature on the topic of filter-generating networks \cite{jia2016nips}, which are deep learning systems of relatively low complexity that generate the synaptic weights in another deep learning system of greater complexity.
The association between the filter-generating network, hereafter denoted as auxiliary network, and the high-complexity network, hereafter denoted as main branch, constitutes an acyclic computation graph named dynamic filter network.
Like any other deep learning system, a dynamic filter network is trained by gradient backpropagation, with both the main branch and the auxiliary branch being updated to minimize the same loss function.
In the computation graph, the two branches merge into a single output branch.
Several mathematical formulations to this merging procedure coexist in the machine learning literature \cite{dai2017iccv,ha2017iclr,li2017cvpr}.
This article compares three of the most straightforward ones, namely adaptive threshold (AT), adaptive weights (AW), and mixture of experts (MoE).

In the application setting of automatic speech recognition, one prominent instance of dynamic filter network is known as context-adaptive neural network (CA-NN) \cite{delcroix2018context}.
In a CA-NN for sound event detection, the purpose of the auxiliary branch is to learn a feature representation that would characterize the intrinsic properties of the acoustic environment, while remaining invariant to whether a sound event is present or not in the environment.
Therefore, the auxiliary branch does not act upon the audio clip itself; but rather, onto some engineered transformation thereof, hereafter known as a vector of \emph{auxiliary features}.

\subsubsection*{Percentile summary statistics as auxiliary features}
Original implementations of CA-NN aim at improving robustness of far-field speech recognition systems to reverberation properties of indoor acoustic environments.
To this effect, they rely on auxiliary features that characterize spatial diffuseness \cite{schwarz2015icassp}, and are derived from a stereophonic audio input.
In contrast, in the application setting of bioacoustic sound event detection, we argue that the leading spurious factor of variability is not reverberation, but rather, background noise.
One distinctive property of background noise, as opposed to foreground events, is that it is locally stationary: although bird calls modulate rapidly in amplitude and frequency, a swarm of insects produce a buzzing noise that remains unchanged at the time scale of several minutes.
Likewise, a vehicle approaching the sensor will typically grow progressively in acoustic intensity, yet without changing much of its short-term spectrum.
We denote by context adaptation (CA) the integration of a sensor-specific, long-term trend into a rapidly changing event detection function, by means of a learned representation of acoustic noise.

It stems from the two observations above that, coarsening the temporal resolution of the time-frequency representation $\mathbf{E}(t,f)$ provides a rough description of the acoustic environment, yet is unaffected by the presence or absence of a short sound event in the short-term vicinity of the time instant $t$.
Hence, we design auxiliary features as nine long-term order statistics (median, quartiles, deciles, percentiles, and permille) summarizing the power spectral density in $\mathbf{E}(t,f)$ over windows of duration $T_\mathrm{CA}$.
In the following, we denote by $\mu(t,q,f)$ the three-way tensor of auxiliary features, where the indices $q$ and $f$ correspond to quantile and mel-frequency respectively.
After cross-validating the parameter $T_\mathrm{CA}$ as a geometric progression ranging between one second and two hours, a preliminary experiment revealed that all values above five minutes led to a background estimator of sufficiently low variance to avoid overfitting.
We set $T_\mathrm{CA}$ to $30$ minutes in the following, and sample $\mu(t,q,f)$ at a rate of $8$ frames per hour.

\subsubsection*{Computational architecture of a context-adaptive neural network}
\begin{figure}
\begin{minipage}{1.0\linewidth}
\centering
\centerline{\includegraphics[width=1.0\linewidth]{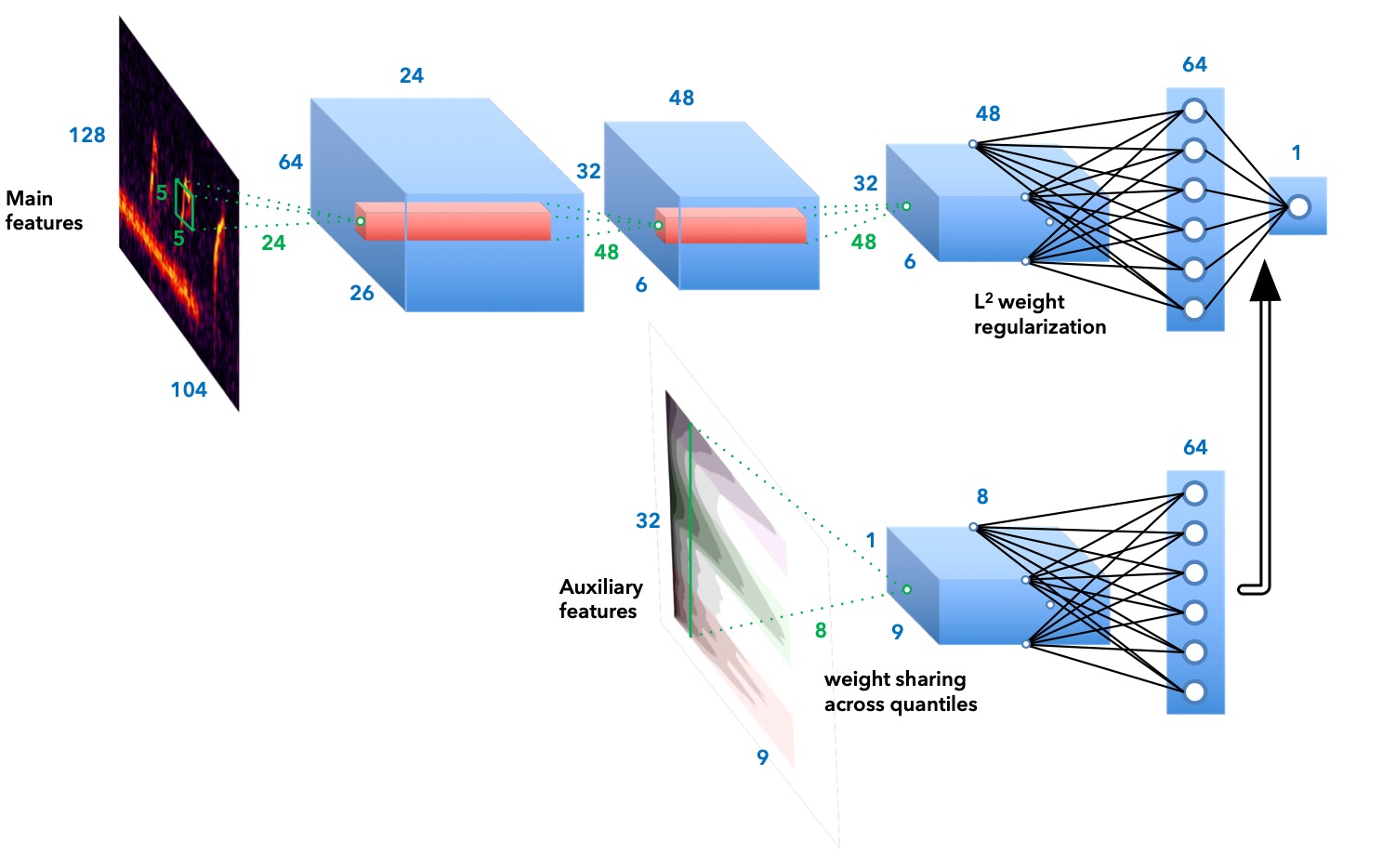}}
\end{minipage}
\caption{Architecture of our context-adaptive neural network (CA-CNN) with spectral summary statistics as auxiliary features.
The double arrow depicts an operation of merging between the main branch and the auxiliary branch.}
\label{fig:ca-cnn-architecture}
\end{figure}

Fig \ref{fig:ca-cnn-architecture} is a block diagram of our proposed context-adaptive neural network (CA-NN) for avian flight call detection in continuous recordings.
The main branch is a convolutional neural network with three convolutional layers followed by two dense layers.
The main branch takes the time-frequency representation $\mathbf{E}(t,f)$ of a short audio clip as input, and learns a $64$-dimensional representation $\mathbf{z}(t,n)$ as output, where $n$ is an integer between $0$ and $63$.
At prediction time, the value taken by $\mathbf{z}(t,n)$ solely depends on the content of the audio clip, and is not context-adaptive.
As regards the auxiliary branch, it is a convolutional neural network with one convolutional layer followed by one dense layer.
The auxiliary branch takes a slice of the tensor of quantile summary statistics $\mu(t,q,f)$ as input, and learns some context-adaptive parameters of arbitrary dimension.
Because the temporal sampling of $\mu$ ($4$ frames per hour) is coarser than the temporal sampling of $y$ ($20$ frames per second), the slice in $\mu(t,q,f)$ that is fed to the network consists of a single temporal frame.
More precisely, it is a matrix of $9$ quantiles $q$ and $32$ mel-frequency bins $f$.

\subsubsection*{Main branch of the context-adaptive neural network}
The main branch has exactly the same architecture as the one that reported state-of-the-art results in urban sound classification \cite{salamon2017spl} (Urban-8K dataset \cite{salamon2014acmmm}) and species classification from clips of avian flight calls \cite{salamon2017icassp} (CLO-43SD dataset \cite{salamon2016plos}).
Its first layer consists of $24$ convolutional kernels of size $5$x$5$, followed by a rectified linear unit (ReLU) and a strided max-pooling operation whose receptive field has a size of $4$x$2$, that is, $4$ logmelspec frames (\ie{} $\SI{6}{\milli\second}$) and $2$ subbands (\ie{} about a musical quartertone).
Likewise, the second layer consists of $24$ convolutional kernels of size $5$x$5$, followed by a ReLU and a strided max-pooling operation whose receptive field has a size of $4$x$2$, that is, $16$ logmelspec frames (\ie{} $\SI{24}{\milli\second}$) and $4$ mel-frequency subbands (\ie{} about a musical semitone).
The third layer consists of $48$ convolutional kernels of size $5\mathrm{x}5$, followed by a ReLU.
There is no pooling after the third layer.
The fourth layer is a fully connected layer with $64$ hidden units, and whose weights are regularized in $L^2$ norm with a multiplicative factor set to $10^{-3}$, followed by a ReLU.
The fifth layer is a fully connected layer with a single output unit, followed by a sigmoid nonlinearity.
We train the whole deep learning architecture to minimize binary cross-entropy by means of the Adam optimizer \cite{kingma2015iclr}.
We use the Keras \cite{chollet2015keras} and pescador \cite{mcfee2017zenodo} Python libraries, respectively to build the model and stream training data efficiently under a fixed memory budget.

\subsubsection*{Auxiliary branch of the context-adaptive neural network}
In the absence of any context adaptation, the last layer of the convolutional neural network for absence \vs{} presence classification of a flight call in the short audio clip $\mathbf{E}(t,f)$ is an affine transformation of the vector $\mathbf{z}$ followed by a sigmoid nonlinearity; that is,
\begin{equation}
    y(t) = \sigma \Big( b + \sum_{n} \mathbf{w}(n) \mathbf{z}(t,n) \Big)
    \label{eq:y-no-context}
\end{equation}
where $\mathbf{w}(n)$ is a $64$-dimensional vector of synaptic weights and the scalar $b$ is a synaptic bias.
Both parameters $\mathbf{w}(n)$ and $b$ are optimized by Adam at training time, yet remain unchanged at prediction time.

The convolutional layer in the auxiliary branch consists of $8$ kernels of size $1$x$32$, followed by a ReLU. 
Observe that, because the height of the kernels is equal to the number of mel-frequency bins in the auxiliary features (\ie{} 32) and does not involve any input padding, this convolutional layer performs weight sharing only across quantiles $q$, and not across neighboring frequency bins $f$.
Each of the learned kernels can be interpreted \emph{post hoc} as a spectral template of background noise, onto which auxiliary features are projected.
The dense layer in the auxiliary branch is an affine transformation from the $9\times8=72$ output activations of the first layer onto $64$ nodes, followed by a ReLU.
In all three cases, we denote by $\mathbf{z}_\mathrm{aux} (t,n)$ the 64-dimensional output of this dense layer.
Because it directly proceeds from the auxiliary features $\mu(t,q,f)$ and not from the main features $\mathbf{E}(t,f)$, $\mathbf{z}_\mathrm{aux}(t,n)$ has a coarse sampling rate of $8$ frames per hour; that is, one context slice every $450$ seconds.
To the best of our knowledge, this paper is the first in proposing to use long-term summary statistics as auxiliary features to a context-adaptive neural network.

In this article, we compare experimentally three formulations of such a feature map: adaptive weights (AW), adaptive threshold (AT), and mixture of experts (MoE).
These formulations correspond to different equations connecting the output $\mathbf{z}(t,n)$ of the main branch with the output $\mathbf{z}_\mathrm{aux}(t,n)$ of the auxiliary branch into a predicted probability of presence $y(t)$ at time $t$, described below.

\subsubsection*{Adaptive weights}
In its adaptive weights formulation (AW), context adaptation replaces $\mathbf{w}(n)$ by $\mathbf{z}_{\mathrm{aux}}(t,n)$ verbatim in Equation \ref{eq:y-no-context}, resulting in an event detection function of the form
\begin{equation}
    y(t) = \sigma \left( b + \sum_{n} \mathbf{z}_{\mathrm{aux}}(t,n) \mathbf{z}(t,n)\right).
\end{equation}

Observe that the CNN baseline is a particular case of this formulation, in which the vector $\mathbf{z}_{\mathrm{aux}}(t,n)$ is constant through time.
This is made possible by setting the synaptic weights of the dense layer in the auxiliary branch to zero, and keeping only nonnegative biases for each of the $64$ nodes.
Therefore, a CA-CNN with adaptive weights has an optimal training accuracy that is, in theory, at least as good as that of a conventional CNN with static weights.
However, because the loss surface of a deep neural network is nonconvex, an iterative stochastic optimizer such as Adam reaches a local optimum rather than the global optimum in the space of neural network parameters.
Consequently, a CA-CNN with adaptive weights may in practice underperform a conventional CNN.

\subsubsection*{Adaptive threshold}
In its adaptive threshold formulation (AT), context adaptation learns a $64$-dimensional static vector $\mathbf{w}_{\mathrm{aux}}(n)$, onto which is projected the auxiliary representation $\mathbf{z}_{\mathrm{aux}}(t,n)$ by canonical inner product.
This inner product replaces the static scalar bias in Equation \ref{eq:y-no-context}, resulting in an event detection function of the form
\begin{equation}
    y(t) = \sigma \left( \sum_{n} \mathbf{w}_{\mathrm{aux}}(n) \mathbf{z}_{\mathrm{aux}}(t,n) + \sum_{n} \mathbf{w}(n) \mathbf{z}(t,n) \right).
\end{equation}

Again, the CNN baseline is a particular case of the AT formulation.
Indeed, setting the vector $\mathbf{z}_{\mathrm{aux}}(t,n)$ to a constant and the weights $\mathbf{w}_{\mathrm{aux}}(n)$ such that the product $\mathbf{w}_{\mathrm{aux}}(n) \mathbf{z}_\mathrm{aux}(t,n)$ is equal to $\mathbf{w}(n)$ for every $n$ is equivalent to discarding context adaptation altogether.

Furthermore, this formulation can also be interpreted as the application of a slowly varying threshold onto a static event detection function.
This is because, by monotonicity of the inverse sigmoid function $\sigma^{-1}$, and given some fixed threshold $\tau$, the inequality $y(t)>\tau$ is equivalent to
\begin{equation}
\sigma \left(\sum_{n} \mathbf{w}(n) \mathbf{z}(t,n) \right) 
>
\sigma\left(\sigma^{-1}(\tau) - \sum_{n} \mathbf{w}_{\mathrm{aux}}(n) \mathbf{z}_{\mathrm{aux}}(t,n)\right).
\end{equation}
The interpretation of the right-hand side as a time-varying threshold is all the more insightful given that $\mathbf{z}_\mathrm{aux}(t)$ has much slower variations than $\mathbf{z}(t)$, \ie{} $8$ frames per hour \vs{} $20$ frames per second.
Under this framework, we may draw a connection between context adaptation in neural networks and a long-lasting line of research on engineering adaptive thresholds for sound onset detection \cite{bello2005tsap}.

\subsubsection*{Mixture of experts}
Under the adaptive weights formulation, each scalar weight in $\mathbf{w}_{\mathrm{aux}}(n)$ is an independent output of the auxiliary network.
In contrast, the mixture of experts formulation (MoE) reduces this requirement by learning a fixed weight vector $\mathbf{w}(n)$ and having a much smaller number of adaptive weights (\eg{} $K=4$) that are applied to subsets of $\mathbf{w}(n)$.
Each of these subsets comprises $\frac{N}{K}$ nodes and can be regarded as an ``expert''.
Therefore, the small number $K$ of outputs from the auxiliary branch no longer corresponds to the number of node weights in the main branch, but to the number of mixture weights across expert subsets, hence the name of ``mixture of experts'' (MoE) formulation \cite{yang2018iclr}.

In practice, the output of the main branch $\mathbf{z}(t,n)$ is reshaped into a tensor $\widetilde{\mathbf{z}}(t,m,k)$, with the integer indices $m$ and $k$ respectively being the quotient and remainder of the Euclidean division of the integer $n$ by the constant $K$.
Likewise, we reshape $\mathbf{w}(t,n)$ into $\widetilde{\mathbf{w}}(t,m,k)$, $\mathbf{z}_{\mathrm{aux}}(t,n)$ into $\widetilde{\mathbf{z}}_{\mathrm{aux}}(t,m,k)$, and $\mathbf{w}_{\mathrm{aux}}(t,n)$ into $\widetilde{\mathbf{w}}_{\mathrm{aux}}(t,m,k)$, where $n=K\times m + k$ for every $0 \leq n < 64$.

The integer $k$, known as expert index, ranges from $0$ to $(K-1)$, and is a hyperparameter of the chose context-adaptive architecture.
In accordance with \cite{delcroix2018context}, we manually set $K=4$ in all of the following.
On the other hand, the integer $m$, known as mixture index, ranges from $0$ to $M=\frac{N}{K}$, \ie{} from $0$ to $M=16$ for $N=64$ nodes and $K=4$ experts.

First, the auxiliary branch converts $\widetilde{\mathbf{z}}_{\mathrm{aux}}(t,m,k)$ into a $K$-dimensional time series $\alpha_{\mathrm{aux}}{t,k}$, by means of an affine transformation over the mixture index $m$:

\begin{equation}
    \alpha_{\mathrm{aux}}(t,k) = b_\mathrm{aux}(k) + \sum_{m} \widetilde{\mathbf{w}}_{\mathrm{aux}}(t,m,k) \widetilde{\mathbf{z}}_{\mathrm{aux}}(t,m,k).
\end{equation}

Secondly, a softmax transformation maps $\alpha_{\mathrm{aux}}(t,k)$ onto a discrete probability distribution over the experts $k$.
Each softmax coefficient then serves as a multiplicative gate to the static inner product between $\widetilde{\mathbf{w}}(t,m,k)$ and $\widetilde{\mathbf{z}}(t,m,k)$ over the mixture index $m$ in the main branch.
This leads to the following definition for the event detection function $y(t)$:
\begin{equation}
    y(t) = \sigma \left( b + \sum_{k} \frac{e^{\alpha_{\mathrm{aux}}(t,k)}}{\sum_{k^\prime} e^{\alpha_{\mathrm{aux}}(t,k^\prime)}} \left( \sum_{m} \widetilde{\mathbf{w}}(t,m,k) \widetilde{\mathbf{z}}(t,m,k) \right) \right).
\end{equation}

Like the AW and AT formulations, the MoE formulation is a generalization of the CNN baseline.
Indeed, setting the learned representation $\widetilde{\mathbf{w}}_{\mathrm{aux}}(t,m,k)$ to zero and the static vector of auxiliary biases $b_{\mathrm{aux}}(k)$ to an arbitrary constant will cause the probability distribution over experts $k$ to be a flat histogram.


\subsection*{Per-channel energy normalization}

\subsubsection*{Definition}

Per-channel energy normalization (PCEN) \cite{wang2017icassp} has recently been proposed as an alternative to the logarithmic transformation of the mel-spectrogram (logmelspec), with the aim of combining dynamic range compression (DRC, also present in logmelspec) and adaptive gain control (AGC) with temporal integration. AGC is a prior stage to DRC involving a low-pass filter $\mathbf{\phi}_T$ of support $T$, thus yielding
\begin{equation}
\mathbf{PCEN}(t,f) =
\left(\dfrac{\mathbf{E}(t,f)}{(\varepsilon+(\mathbf{E}\overset{t}{\ast}\boldsymbol{\phi}_T)(t,f))^\alpha} + \delta\right)^r - \delta^r
\label{eq:pcen}
\end{equation}
where $\alpha, \varepsilon, \delta$, and $r$ are positive constants.
While DRC reduces the variance of foreground loudness, AGC is intended to suppress stationary background noise.
The resulting representation has shown to improve performance in far-field ASR \cite{battenberg2017arxiv}, keyword spotting \cite{wang2017icassp}, and speech-to-text systems \cite{shan2018attention}.

There is practical evidence that, over a large class of real-world recording conditions, PCEN decorrelates and Gaussianizes the background while enhancing the contrast between the foreground and the background \cite{lostanlen2018spl}.
From the standpoint of machine learning theory, this Gaussianization property appears to play a key role in avoiding statistical overfitting.
Indeed, deep neural networks are optimally robust to adversarial additive perturbations if the background in the training set is a realization of additive, white, and Gaussian noise (AWGN) \cite{franceschi2018aistats}.
This theoretical argument is all the more crucial to the success of sound event detection systems given that, in the case of bioacoustic sensor networks, background noise is nonuniform, and thus will typically vary in terms of power spectral density between training set and test set.
Therefore, in our study, PCEN serves the double purpose of, first, disentangling foreground and background as independent sources of variability and, second, facilitating the transferability of learned audio representations between one recording condition and another.

\subsubsection*{Parameter settings}

As Equation \ref{eq:pcen} shows, the instantiation of PCEN depends upon six parameters: $T$, $\alpha$, $\varepsilon$, $\delta$, and $r$.
The effect of these parameters relate to respective properties of the foreground and background noise, as well as the underlying choice of time-frequency representation $\mathbf{E}(t,f)$.
Yet, the motivation for developing PCEN initially arose in the context of far-field automatic speech recognition in domestic environments \cite{krstulovic2018casse}.
In this context, the recommendations of the original publication on PCEN \cite{wang2017icassp} are as follows: $\varepsilon = 10^{-6}$; $\alpha = 0.98$ ; $\boldsymbol{\delta} = 2$; $r = \frac{1}{2}$; and $T_{\mathrm{PCEN}} = \SI{400}{\milli\second}$.

\begin{figure}
\begin{minipage}{0.95\linewidth}
\centering
\centerline{\includegraphics[width=1.0\linewidth]{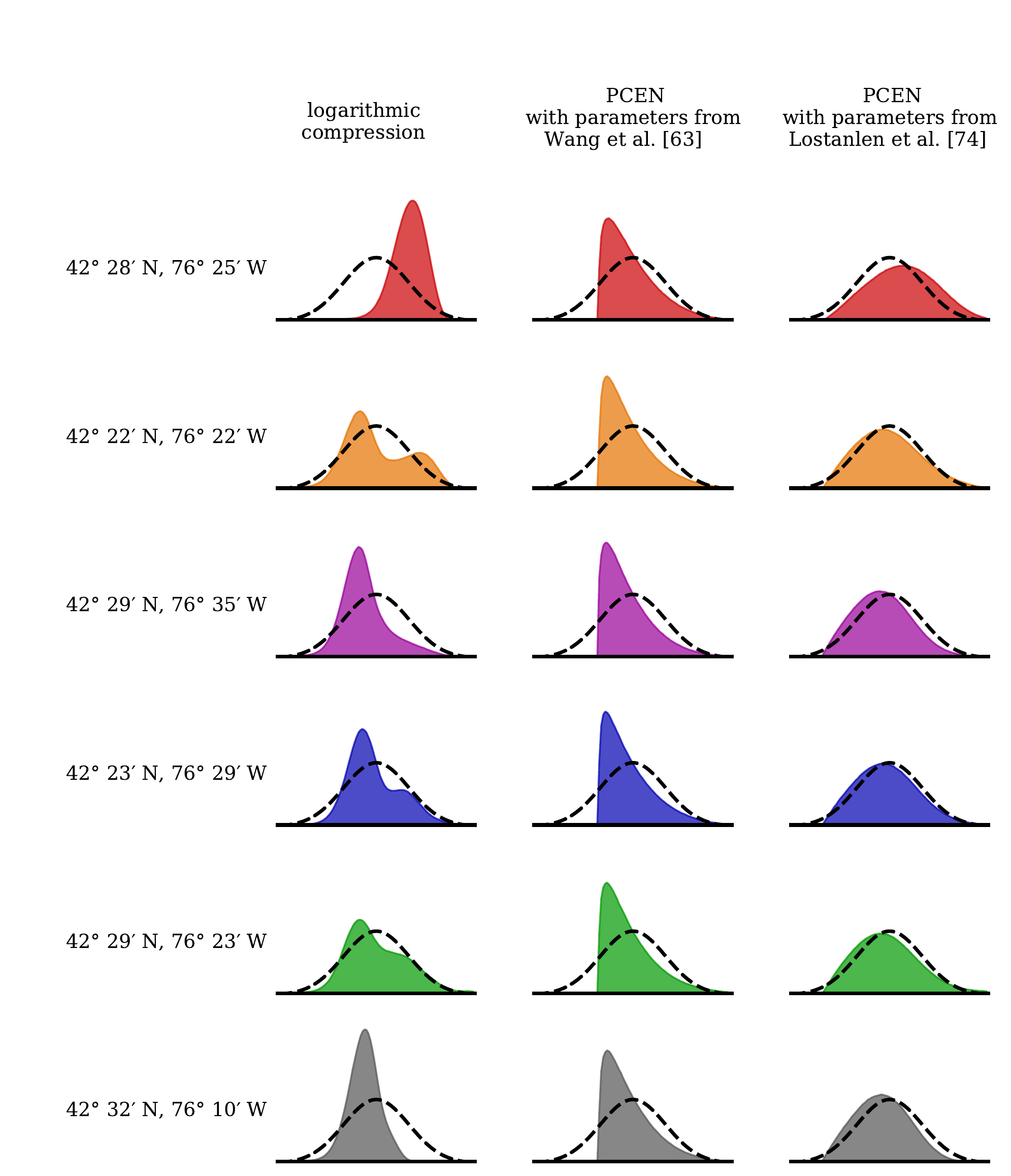}}
\end{minipage}
\caption{Histogram of magnitudes of time--frequency bins for six one-minute recordings in the BirdVox-full-night dataset, corresponding to six different bioacoustic sensors, for different choices of loudness mapping. Left: after logarithmic compression. Middle: after per-channel energy normalization (PCEN) with the parameters of \cite{wang2017icassp}, best suited to automatic speech recognition in a noisy indoor environment. Right: after PCEN with the parameters of \cite{lostanlen2018spl}, best suited to sound event detection in a noisy outdoor environment. We globally standardize the magnitudes in each column to null mean and unit variance. The black dashed line represents the ideal case of a standard normal distribution. The text on the left of each row denotes the latitude and longitude of the corresponding autonomous recording unit in the acoustic sensor network. See text for details about parameter settings.}
\label{fig:pcen-gaussianization}
\end{figure}

In contrast, the detection of avian flight calls in rural outdoor areas with autonomous recording units is a starkly different application setting, thus requiring adjustments in the choice of parameters.
One previous publication \cite{lostanlen2018spl} has conducted an asymptotic analysis of PCEN components, and concluded with some practical recommendation for making such adjustments according to the task at hand.
It appears that, in comparison with indoor applications (\eg{} ASR in the smart home), bioacoustic event detection distinguishes itself by faster modulations of foreground, higher skewness of background magnitudes, a louder background, and more distant sources.
Such idiosyncrasies respectively call for a lower $T$, a lower $\alpha$, a higher $\delta$, and a lower $r$.

We decode each audio signal as a sequence of floating-point numbers in the range $[-2^{31}; 2^{31}[$ with a sample rate of \SI{22,050}{\hertz}, and apply a short-term Fourier transform (STFT) with window size $256$ (\SI{12}{\milli\second}), hop size $32$ (\SI{1.5}{\milli\second}), and fast Fourier transform length ($N_{\mathrm{FFT}}$ parameter) $1024$.
Then, we map the $513$ nonnegative frequency bins of the STFT squared modulus representation onto a mel scale, with $128$ mel-frequency subbands ranging from \SI{2}{\kilo\hertz} to \SI{11,025}{\kilo\hertz}.
Lastly, we apply PCEN according to Equation \ref{eq:pcen} with $\varepsilon = 10^{-6}$; $\alpha = 0.8$ ; $\boldsymbol{\delta} = 10$; $r = \frac{1}{4}$; and $T_{\mathrm{PCEN}} = \SI{60}{\milli\second}$ after following the recommendations of \cite{lostanlen2018spl}.
The choice of minimal frequency at $\SI{2}{\kilo\hertz}$ corresponds to a lower bound on the vocal range of avian flight calls of thrushes.
With the librosa Python library \cite{brian_mcfee_2018_1252297}, the computation of logmelspec is about $20$ times faster than real time on a dual-core Intel Xeon E-2690v2 $\SI{3.0}{\giga\hertz}$ central processing unit (CPU).
Replacing these ad hoc constants by trainable, frequency-dependent parameters $\boldsymbol{\alpha}(f)$, $\boldsymbol{\delta}(f)$, and so forth, is a promising line of research \cite{zinemanas2019fruct,millet2019icassp}, but is beyond the scope of this paper, as it does not fundamentally change its overall narrative.

Fig \ref{fig:pcen-gaussianization} illustrates the effect of PCEN on the empirical distribution of mel-frequency spectrogram magnitudes, across all six sensor locations in the BirdVox-full-night dataset.
First, logarithmic compression (left column) results in non-Gaussian, occasionally bimodal distributions whose mode may vary from one sensor to the next, by as much as one global standard deviation.
This fact demonstrates that logmelspec representations are fundamentally inadequate for deploying a single machine listening system over multiple autonomous recording units.
Secondly, running PCEN with parameters that are best suited to indoor environments, such as those of \cite{wang2017icassp} (middle), results in distributions of magnitudes that are consistent across outdoor recording locations, but are skewed towards the right.
This fact demonstrates the importance of adjusting PCEN parameters, at least roughly, to the type of acoustic sensor network (indoor \vs outdoor).
Lastly, running PCEN with parameters that are best suited to outdoor environments, such as those recommended by \cite{lostanlen2018spl} (right), results in distributions that are quasi-Gaussian, consistently across sensors.

\subsection*{Baseline: convolutional neural network}

The baseline model of our study is a CNN in the logmelspec domain for avian flight call detection, whose architecture is replicated from a previous study \cite{lostanlen2017icassp}.
In spite of its simplicity, this deep learning model has shown to significantly outperform other algorithms for avian flight call detection in the BirdVox-full-night dataset, including spectral flux \cite{bello2005tsap}, the Vesper library reimplementation of the ``Old Bird'' energy-based detection function \cite{mills2018zenodo}, and the PCA-SKM-SVM shallow learning pipeline \cite{salamon2017spl}.

In a preliminary stage, we explored over $100$ common variations in the architecture of the baseline, including changes in kernel size, layer width, number of layers, mel scale discretization, multiresolution input \cite{anden2015mlsp}, choice of nonlinearity, use of dropout, use of batch normalization, and choice of learning rate schedule.
Yet, none of these general-purpose variations, unrelated in their design to the question of robustness to background noise, led to systematic improvements upon the baseline.
Therefore, although the baseline architecture is by no means optimal, there are grounds to believe that the following improvements brought by CA and PCEN would not easily be matched by applying other, more well-established variations.

\subsection*{Artificial data augmentation}

Applying randomized digital audio effects to every sample in a dataset at training time often reduces overfitting without any extra computational cost at prediction time \cite{mcfee2015ismir}.
Hence, many deep machine listening systems are trained on augmented data: related applications to this study include bird species classification \cite{salamon2017icassp}, singing voice detection \cite{schluter2015ismir}, and urban sound classification \cite{salamon2017spl}.
Yet, one difficulty of artificial data augmentation is that the chosen distribution of parameters needs to reflect the underlying variability of the data.
In the context of avian flight calls, we use domain-specific knowledge in animal behavior so as to find an appropriate range of parameters for each perturbation.

We distinguish two kinds of data augmentation: geometrical and adaptive.
Geometrical data augmentation (GDA) includes all digital audio effects whose parameters are independent of the probability distribution of samples in the training set, such as pitch shifting and time stretching.
On the contrary, adaptive data augmentation (ADA) takes into account the whole training data, and in some cases also the corresponding labels, to transform each sample.
For instance, mixing each audio clip in the sensor at hand with a negative (noisy) audio clip belonging to a different sensor in the training set leads to greater generalizability \cite{salamon2017waspaa}.
However, this adaptive procedure causes the number of augmented samples to scale quadratically with the number of sensors.
Furthermore, it cannot be easily combined with CA because the addition of extraneous noise to the front end would require to also re-compute the corresponding auxiliary features for the mixture of signal and noise at a long temporal scale ($T_\mathrm{CA}=30$ minutes), which is intractable for large $T_\mathrm{CA}$.
Therefore, we apply geometrical data augmentation to all models, but adaptive data augmentation only to the models that do not include context adaptation.

We use the \textsf{muda} Python package (MUsical Data Augmentation \cite{mcfee2015ismir}) to apply $20$ randomized digital audio effects to each audio clip: four pitch shifts; four time stretchings; and four additions of background noise originating from each of all three training sensors in the cross-validation fold at hand.
The choices of probability distributions and hyperparameters underlying these augmentations are identical to those of \cite{lostanlen2017icassp}, and are chosen in accordance with expert knowledge about the typical vocal ranges of thrushes, warblers, and sparrows.

All transformations are independent from each other in the probabilistic sense, and never applied in combination.
In the case of the addition of background noise, we restrict the set of augmentations to those in which the background noise and the original audio clip belong to recordings that are both in the training set, or both in the validation set.

\section*{Experimental design}

As illustrated in Fig \ref{fig:binary-classification-diagram}, our experimental design consists of two stages: training the system as a binary classifier of presence \vs{} absence, and then running it on a task of avian flight call detection.
The datasets we use for these two stages are BirdVox-70k and BirdVox-full-night.
The metrics we use are miss rate (\ie{}, the complementary of classification accuracy) and Area Under the Precision--Recall Curve (AUPRC).
While the former is related to the way our deep learning models are trained, the latter is related to the way these models are deployed in the field.

\subsection*{Stage 1: binary classifier}

\subsubsection*{Dataset: BirdVox-70k}

We train all sound event detection models presented in this article as binary classifiers of presence \vs{} absence of a flight call, at the time scale of audio clips of duration $\SI{150}{\milli\second}$.
To this end, we rely on the BirdVox-$70$k dataset, which contains $35$k positive clips and $35$k negative clips, originating from a network of $6$ bioacoustic sensors.
We refer to \cite{lostanlen2017icassp} for more details on the curation of the BirdVox-$70$k dataset.

\subsubsection*{Evaluation: leave-one-sensor-out cross-validation}

Because our study focuses on the comparative generalizability of automated systems for flight call detection, we split the BirdVox-$70$k dataset according to a stratified, ``leave-one-sensor-out'' evaluation procedure.
After training all systems on the audio recordings originating from three sensors (training set), we use two of the remaining sensors to identify the optimal combination of hyperparameters (validation set), and leave the last sensor out for reporting final results (test set).
From one fold to the next, all boundaries between subsets shift by one sensor, in a periodic fashion.

The loss function for training the system is binary cross-entropy $\mathcal{L}(y,y_{\mathrm{true}}) = \log \vert y - y_{\mathrm{true}} \vert$, where $y_{\mathrm{true}}$ is set to $1$ if a flight call is present in the audio clip at hand, and $0$ otherwise.
To evaluate the system in its validation stage, we measure a classification accuracy metric; that is, the proportion of clips in which the absolute difference $\vert y - y_{\mathrm{true}}\vert$ is below $0.5$ over a hold-out validation set.

\subsection*{Stage 2: detection in continuous audio}

\subsubsection*{Dataset: BirdVox-full-night}
We evaluate all sound event detection models presented in this article on a task of species-agnostic avian flight call detection.
To this end, we rely on the BirdVox-full-night dataset, which contains recordings originating from one ten-hour night of fall migration, as recorded from $6$ different sensors.
Each of these sensors is located in rural areas near Ithaca, NY, USA, and is equipped with one omnidirectional microphone of moderate cost.
The resulting bioacoustic sensor network covers a total land area of approximately $\SI{1000}{\kilo\meter^2}$.
The $6$ recordings in BirdVox-full-night amount to $62$ hours of monaural audio data.
The split between training set, validation set, and test set follows the same ``leave-one-sensor-out'' evaluation procedure as presented in the previous section.
Therefore, all models are tested on recording conditions that are extraneous to the training and validation subsets.
We refer to \cite{salamon2016plos} for more details on sensor hardware and to \cite{lostanlen2017icassp} for more details on BirdVox-full-night.

\subsubsection*{Evaluation: precision and recall metrics}

We formulate the task of avian flight call detection as follows: given a continuous audio recording from dusk to dawn, the system should produce a list of timestamps, each of them denoting the temporal center of a different flight call.
Then, we may evaluate the effectiveness of the system by comparing this list of timestamps against an expert annotation.
To this aim, we begin by extracting local peaks in the event detection function according to a fixed threshold $\tau$.
The baseline CNN model of \cite{lostanlen2017icassp} constrains consecutive detections to be spaced in time by a minimum lag of at least
$\SI{150}{\milli\second}$.
This constraint improved precision in the baseline CNN model without much detriment to recall, and for consistency we kept this constraint throughout our evaluation. However, as we will see in the Results section, this constraint becomes unnecessary in our state-of-the-art combined model, named BirdVoxDetect.
Therefore, BirdVoxDetect may produce predicted timestamps as close to each other as $\SI{100}{\milli\second}$ (\ie{} two discrete hops of duration $\SI{50}{\milli\second}$) as it does not induce any constraint on the minimum duration between adjacent peaks in the event detection function.

Once the procedure of thresholding and peak-picking is complete, the detected peaks (flight calls) are evaluated by matching them to the manually labeled calls --- the ``reference'', sometimes called ``ground truth'' --- and computing the number of true positives (TP), false positives (FP) and false negatives (FN).
This process is repeated for varying peak detection threshold values $\tau$ between 0 and 1 to obtain the standard information retrieval metric of Area Under the Precision--Recall Curve (AUPRC) with 0 being the worst value and 1 being the best.
A detected peak and a reference peak are considered to be a matching pair if they are within $\SI{500}{\milli\second}$ of each other.
To ensure optimum matching of detected peaks to reference peaks while ensuring each reference peak can only be matched to a single estimated peak, we treat the problem as a maximum bipartite graph matching problem \cite{hopcroft1973siam} and use the implementation provided in the \texttt{mir\_eval} Python library for efficiency and transparency \cite{raffel2014ismir}.

\section*{Results}

\subsection*{Stage 1: training of a binary classifier}

\subsubsection*{Exhaustive benchmark on validation set}

\begin{figure}
\begin{minipage}{1.0\linewidth}
\centering
\centerline{\includegraphics[width=1.0\linewidth]{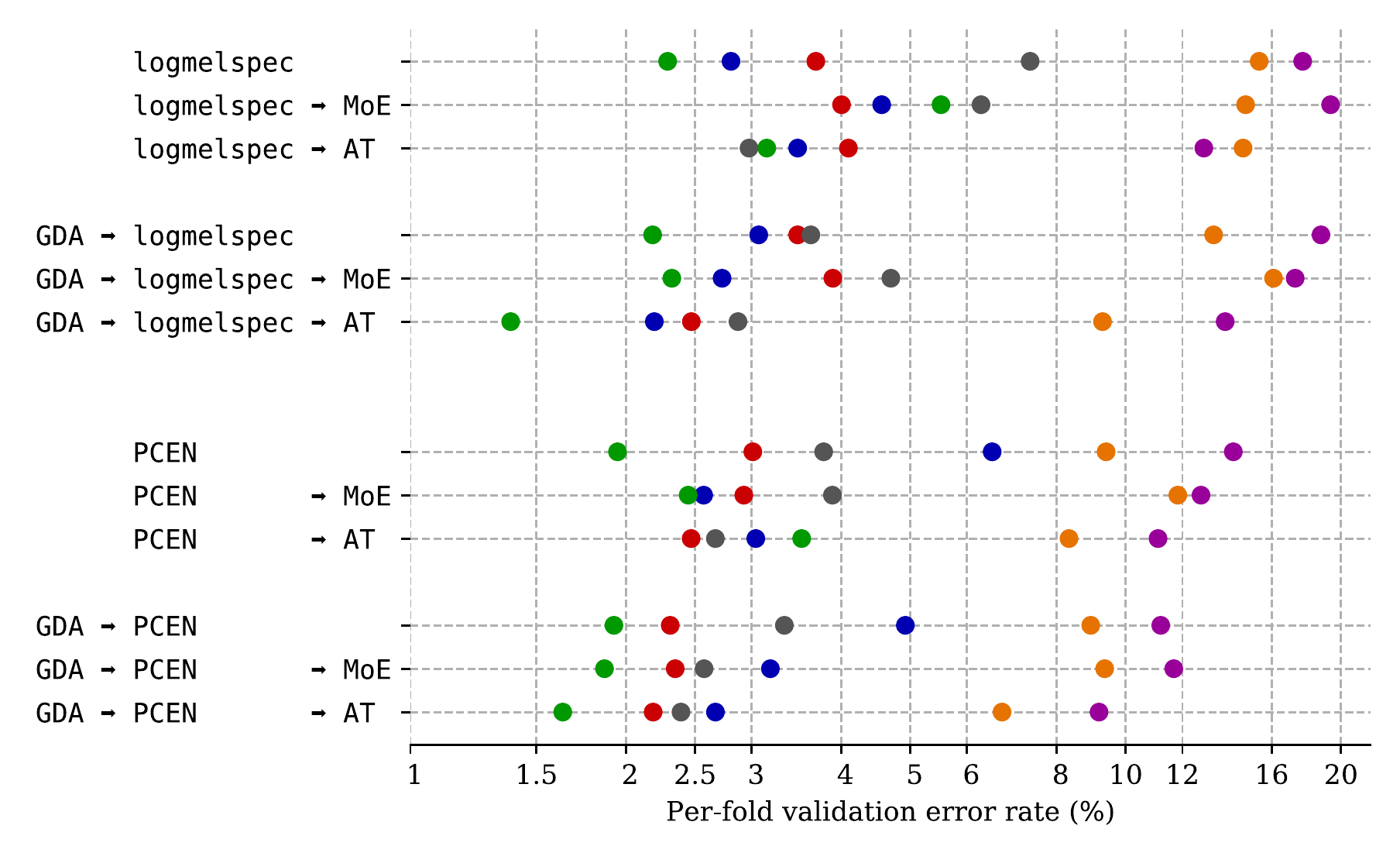}}
\end{minipage}
\caption{Exhaustive benchmark of architectural variations with respect to the logmelspec-CNN baseline, on a task of binary classification of presence \vs{} absence of a bird in audio clips. Dot colors represent folds in BirdVox-$70$k. GDA: geometrical data augmentation. logmelspec: log-mel-spectrogram. PCEN: per-channel energy normalization. MoE: mixture of experts. AT: adaptive threshold.\label{fig:exhaustive-benchmark}}
\end{figure}

Fig \ref{fig:exhaustive-benchmark} summarizes the validation error rates on BirdVox-$70$k of twelve different models.
These models represent different combinations between three design choices: choice of time-frequency representation, choice of formulation in the context adaptation, and use of artificial data augmentation.
In order to mitigate the influence of random initialization on these validation error rates, we train and evaluate each of the twelve different models ten different times on each of the six folds, and report the median validation error rate only.
The cumulative computational budget for training all models is of the order of $180$ GPU-days for training, and $180$ CPU-days for prediction.
In both cases, we parallelize massively across models, folds, and trials, resulting in $720$ different jobs in total, each running independently for approximately six hours on a high-performance computing cluster.

We find that, across all models, some folds consistently lead to a greater error rate than others.
In the case of the logmelspec-CNN baseline, the typical error rate is of the order of $5\%$, but varies between $2\%$ and $20\%$ between folds.
Individual variations of that baseline are not equally beneficial.
First, replacing the logmelspec acoustic frontend by PCEN improves validation accuracy on five folds out of six, and GDA improves it on four folds out of six.
Secondly, adding context adaptation to the baseline, by means of a mixture of experts (MoE), is detrimental to validation accuracy in four folds, while using an adaptive threshold (AT) instead of MoE essentially leaves the baseline unchanged, as it improves and degrades per-fold performance in comparable measures.
Therefore, it appears that context adaptation alone fails to improve the generalizability of a logmelspec-based deep learning model for avian flight call detection.
In what follows, we focus on analyzing the effects of context adaptation on models that are either trained with PCEN, GDA, or both.

Fig \ref{fig:exhaustive-benchmark} also shows that applying GDA to a PCEN-based model consistently improves validation accuracy over all six folds, whether an AT is present or not.
We hypothesize that this consistent improvement is the relational effect of artificial pitch shifts in GDA and background noise reduction caused by PCEN. 
Indeed, one shortcoming of pitch shifting in GDA is that it affects foreground and background simultaneously.
Yet, natural factors of variability in avian flight call detection, such as those arising due to animal behavior, will typically affect the absolute fundamental frequency of the foreground while leaving the background --- \ie{} the power spectral density of insects or passing cars --- unchanged.
Consequently, artificial pitch shifts, even as small as a musical semitone, may lead to a plausible foreground, yet mixed with an implausible background in logmelspec domain.
On the contrary, as described in the Methods section, PCEN tends to bring the distribution of background time-frequency magnitudes closer to additive white Gaussian noise (AWGN).
Because AWGN has a flat spectrum, transposing a polyphonic mixture containing a nonstationary foreground and an AWGN background has the same effect as transposing the foreground only while leaving the background unchanged.
Therefore, not only does replacing the logmelspec acoustic frontend by PCEN improve the robustness of a classifier to background noise, it also helps disentangling pitch transpositions of background and foreground, thus allowing for more extensive geometrical data augmentation by pitch shifts and time stretchings.

Because the six folds in the leave-one-sensor-out cross-validation procedure are of unequal size and acoustic diversity, it is not straightforward to rank all twelve models according to a single global evaluation metric.
However, we may induce a structure of partial ordering between models by the following definition: a model A is regarded as superior to model B if and only if switching from A to B degrades accuracy on half of the folds or more.
According to this definition, the last model (GDA-PCEN-AT) is the only one that is superior to all others.
Moreover, we find that PCEN is superior to the logmelspec baseline; that GDA-PCEN is superior to PCEN; and that GDA-PCEN-AT is superior to PCEN.
We also find that GDA-logmelspec is superior to logmelspec, and that GDA-PCEN is superior to GDA-logmelspec.
However, we do not find either logmelspec-AT or logmelspec-MoE to be superior to logmelspec.
In addition, GDA-PCEN-MoE is superior to GDA-PCEN, yet inferior to GDA-PCEN-AT.

Because GDA-PCEN-AT and GDA-PCEN-MoE perform almost equally across the board, one supplementary question that arises from this benchmark is whether AT and MoE could somehow be combined into a hybrid form of context adaptation.
To challenge this hypothesis, we trained a thirteenth model, named GDA-PCEN-AT-MoE, ten times on each fold of BirdVox-$70$k, and measure median validation accuracies.
We found that this model performs below GDA-PCEN-AT on the majority of folds, and failed to train at all on many trials.
Therefore, we do not pursue this line of research further.
Rather, we adopt the adaptive threshold (AT) formulation as a simple, yet effective, method.
We postulate that the overall degradation in accuracy from GDA-PCEN-AT to GDA-PCEN-AT-MoE is caused by an excessive number of degrees of freedom in the design of the context-adaptive neural network.

Two conclusions arise from all the observations above.
First, the best performing model, in terms of validation accuracy on BirdVox-$70$k, appears to be GDA-PCEN-AT.
Therefore, in the following, GDA-PCEN-AT is the model that we will choose to report results on the test set.
Second, because context adaptation does not improve the baseline, but only models that feature PCEN, we deduce that an ablation study from GDA-PCEN-AT should begin by removing AT before removing PCEN.
Therefore, in the following, we discuss and compare the evolution of test set recall through time for GDA-PCEN-AT and GDA-PCEN, but do not report test set results on GDA-logmelspec-AT because this model is excluded by cross-validation.

As a supplement to this benchmark, we evaluate a completely different deep learning system, named ``bulbul'', for the detection of bird sounds in audio signals \cite{grill2017eusipco}.
This system won the 2017 edition of the ``Bird detection in audio'' challenge \cite{stowell2016mlsp}, and its source code is freely available at the following URL address: \url{https://jobim.ofai.at/gitlab/gr/bird_audio_detection_challenge_2017}.
With the help of the authors, we set up a pre-trained version of bulbul and use it as a binary classifier of presence \vs{} absence in the BirdVox-70k dataset.
Given that it expects 10-second audio clips rather than 150-millisecond audio clips, we had to repeat each clip periodically 67 times in order to collect each prediction. 
After calibraing the detection threshold to its optimal value, we report a detection accuracy of $50.8\%$, that is, marginally above the chance level at $50\%$.
In comparison, the deep learning baseline of \cite{lostanlen2018spl} has a detection accuracy of $90.5\%$.
This result confirms that the deep learning baseline of \cite{lostanlen2018spl} was the state of the art in avian flight call detection up until this publication.
It also shows, as stated in the introduction, that the task of avian flight call detection at the fine time scale of isolated acoustic events is fundamentally different from the task of ``bird detection in audio'', as formulated by \cite{stowell2018mee} at the scale of ten seconds; and that state-of-the-art systems in the coarse-scale task are ill-suited to address the fine-scale task.

\subsubsection*{Ablation study}
\begin{figure}
\centering
\centerline{\includegraphics[width=1.0\linewidth]{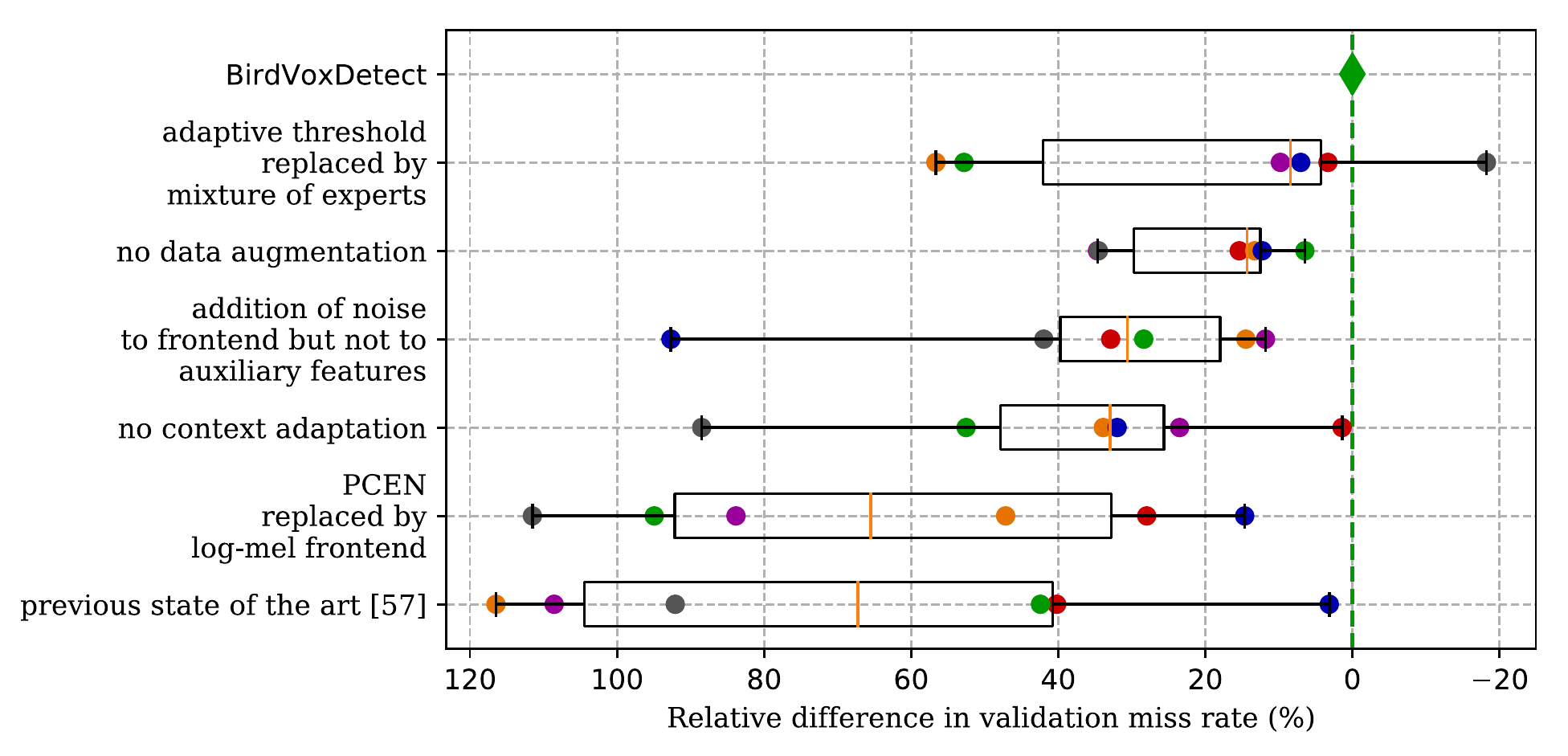}}
\caption{Ablation study of best model (CNN+PCEN+CA+GDA) on the BirdVox-full-night dataset. Boxes (\resp{} whiskers) denote interquartile (\resp{} extremal) variations between sensors.}
\label{fig:ablation-study}
\end{figure}

Once the exhaustive benchmark has identified one reference model --- namely, GDA-PCEN-AT --- we may measure the relative difference in error rate between the reference and some other model in the benchmark for each fold, and compute quantiles across folds.
The reason why we opt for averaging relative differences rather than absolute differences is that the former, unlike the latter, tends to follow a symmetric distribution across folds, and thus can be represented on a box-and-whisker plot.
Thus, we may compare and rank the respective positions of the boxes for different ablations of the reference.
Furtheremore, relative improvements, unlike absolute improvements, are theoretically unbounded.
Fig \ref{fig:ablation-study} summarizes the results of our ablation study.

First, replacing AT by MoE hardly affects our results.
Therefore, it is more likely the presence of any form of context adaptation at all, rather than specific architectural choices in the side-channel neural network, that enables a greater generalization across folds.
In the original article on context-adaptive neural networks \cite{delcroix2015icassp}, as well as those which followed by the same first author \cite{delcroix2016icassp,delcroix2016interspeech,delcroix2018context}, the proposed technique is MoE.
Although we confirm that MoE is successful, we find that implementing context adaptation with an adaptive threshold (AT) leads to sound event detection results which are within a statistical tie with respect to MoE. 
Yet, AT is simpler, faster, and more interpretable than MoE.
Therefore, although our alternative techniques for context adaptation do not lead to significant improvements in accuracy, our results call into question the widespread claim that the root cause behind the success of context adaptation in deep learning lies in its ability to represent multiplicative interactions between heterogeneous sources of data, and thus elicit neural specialization at prediction time.
Although the MoE and AW techniques do contain multiplicative interactions, the AT technique does not. It stems from these observations that, at least in the case of long-term bioacoustic monitoring with spectrotemporal summary statistics as auxiliary features, the presence or absence of multiplicative interactions is not the sole determining factor of the success of context adaptation. 

Secondly, removing geometrical data augmentation, and training the PCEN-AT model on original audio clips from BirdVox-$70$k only, does hinder accuracy consistently, though less so than other improvements upon the baseline (i.e. PCEN and context adaptation).
This supports our hypothesis that shortcomings of the baseline are mainly attributable to its lack of robustness to background noise, more so than its lack of robustness to the geometrical variability in time-frequency patterns of avian flight calls.

Thirdly, we find that ADA-PCEN-AT and GDA-PCEN bring comparable differences in miss rate with respect to the reference model GDA-PCEN-AT.
In other words, the addition of noise to the main branch of the network without reflecting it in the auxiliary features is, quite unsurprisingly, about as detrimental as not having auxiliary features at all.

Fourthly, replacing PCEN by logmelspec in the reference model increases miss rates by about $60\%$ on average, and over $100\%$ in two out of the six folds.
Thus, there are grounds to believe that PCEN is the predominant contributor to validation accuracy in the GDA-PCEN-AT reference model.

\subsection*{Stage 2: detection in continuous audio}

\subsubsection*{Precision-recall curves}
Although the BirdVox-$70$k dataset is particularly well suited for training machine listening systems for avian flight call detection, it does not reflect the practical use case of flight call monitoring in continuous recordings.
Indeed, as described in \cite{lostanlen2017icassp}, BirdVox-$70$k is curated in a semi-automatic fashion: while the positive clips proceed from human annotations, the negative clips correspond to the false alarms of an off-the-shelf shallow learning model.
Specifically, BirdVox-$70$k contains a larger proportion of challenging confounding factors --- such as siren horns and electronic beeps --- and, conversely, a smaller proportion of quasi-silent sound clips, than a full night of bird migration.
Therefore, whereas the previous subsection used validation accuracy on BirdVox-$70$k as a proxy for singling out an optimal model, it is, from the perspective of applied bioacoustics, less insightful to report test set accuracy on BirdVox-$70$k than it is to plot a precision-recall curve on BirdVox-full-night.

\begin{figure}
\centering
\centerline{\includegraphics[width=0.8\linewidth]{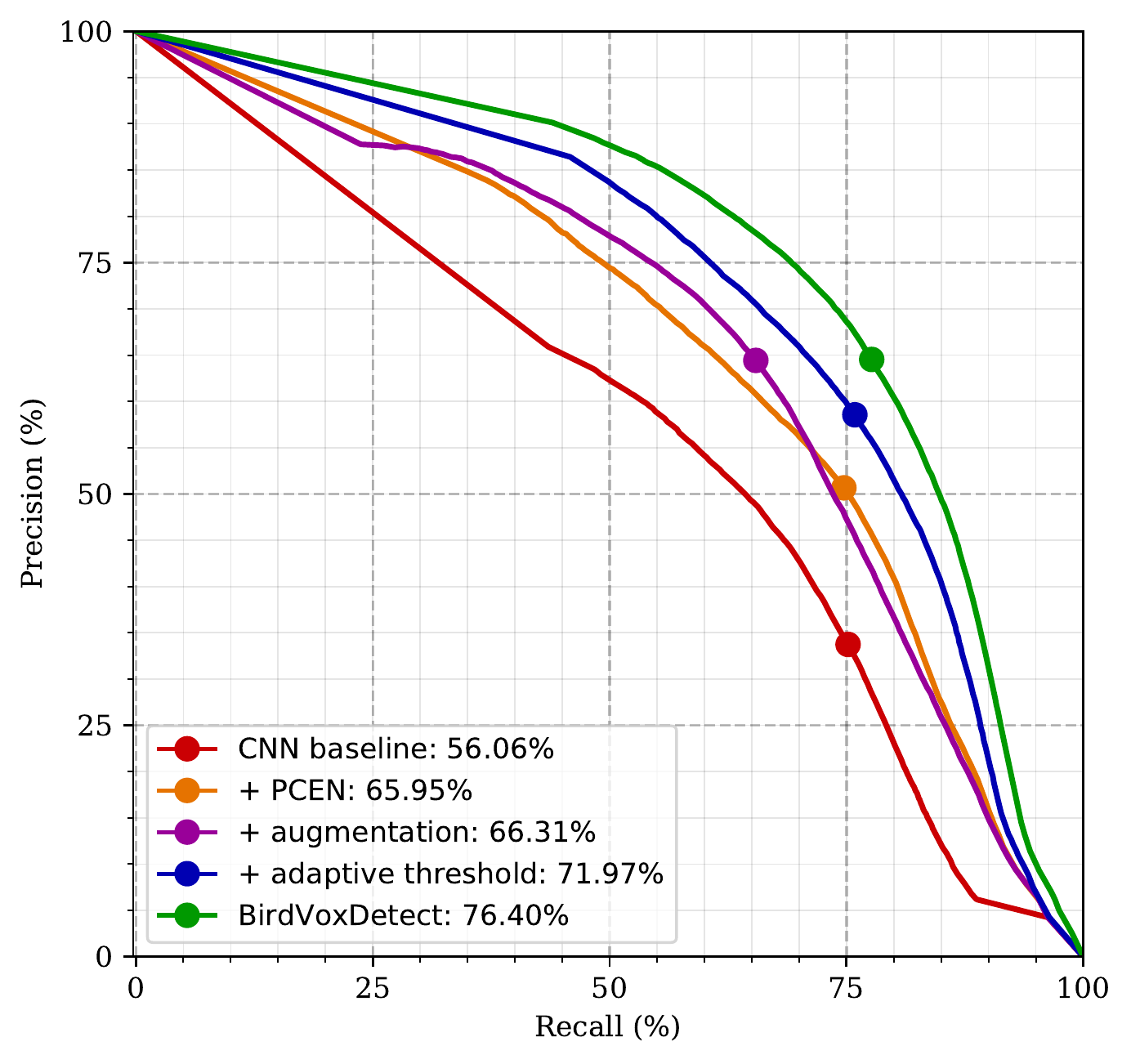}}
\caption{Precision-recall detection curves in avian flight call detection (BirdVox-full-night dataset). The area under each precision-recall curve (AUPRC) is shown in the legend of the plot. The red line ($56.06\%$) is the CNN baseline (previous state of the art of \cite{lostanlen2017icassp} without context adaptation). The blue line ($76.40\%$) is our best performing model on the validation set, and is released under the name of BirdVoxDetect. The thick dot on each curve denotes the optimal tradeoff between precision and recall, corresponding to a maximal \Fscore{}. CNN: convolutional neural network; PCEN: per-channel energy normalization.}
\label{fig:precision-recall}
\end{figure}

Fig \ref{fig:precision-recall} illustrates the combined effects of PCEN, CA, and GDA on the area under the precision-recall curve (AUPRC) when applying CNN to flight call event detection on the BirdVox-full-night test set recordings.
In agreement with the ablation study, the best model (CNN+PCEN+CA+GDA) reaches a test AUPRC of $72.0\%$, thus outperforming models lacking either PCEN, CA, or GDA.
In addition to the precision-recall curves that are shown in Fig \ref{fig:precision-recall}, we computed predictions over BirdVox-full-night for each of the twelve models presented in the exhaustive benchmark (Fig \ref{fig:exhaustive-benchmark}), over $6$ folds and $10$ randomized trials.
This last procedure represents about $10$ CPU-years of computation in total.
From it, we can confirm that BirdVoxDetect does not overfit the validation set more than any of its counterparts.

\subsubsection*{Error analysis}

\begin{figure}
\centering
\centerline{\includegraphics[width=0.75\linewidth]{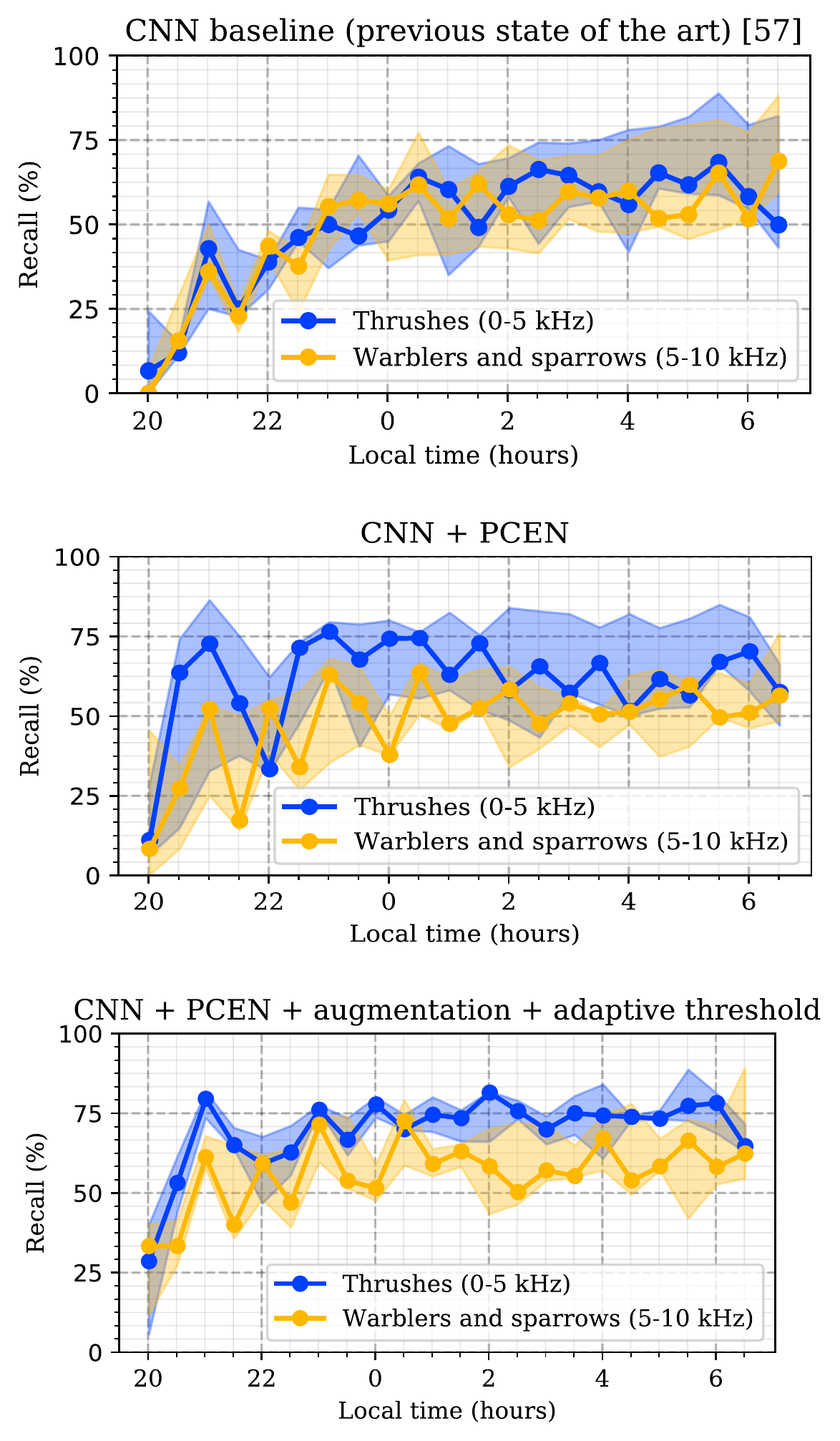}}
\caption{Evolution of recall in the automatic detection of avian flight calls over $30$-minute segments in BirdVox-full-night, for two taxa of migratory birds: thrushes (blue curve, 0-\SI{5}{\kilo\hertz} frequency range) and warblers and sparrows (orange curve, 5-\SI{10}{\kilo\hertz} frequency range).
CNN: convolutional neural network.
PCEN: per-channel energy normalization.
Shaded areas denote interquartile variations across sensors.
We find that PCEN improves robustness to noise nonstationarity, while context adaptation improves robustness to noise nonuniformity.}
\label{fig:nonstationarity-vs-nonuniformity}
\end{figure}

We opened this article by pointing out that many state-of-the-art systems for bioacoustic event detection lack robustness to spatiotemporal variations in background noise, thus preventing their reliability at the scale of distributed sensor networks.
In particular, we had shown in Fig \ref{fig:icassp-sota} that the CNN baseline of \cite{lostanlen2017icassp} exhibits a poor recall (below 50\%) in the early hours of BirdVox-full-night, while only achieving a satisfying recall towards the end of each full night continuous recording.
We hypothesized that such drastic variation in performance was attributable to the scarcity of training examples at dusk in comparison to dawn, in conjunction with more intense levels of background noise at dusk than at dawn.
Now, Fig \ref{fig:nonstationarity-vs-nonuniformity} offers evidence to support this initial hypothesis.
While the top subfigure shows the evolution of recall of the CNN baseline on the BirdVox-full-night dataset, the other two subfigures in Fig \ref{fig:nonstationarity-vs-nonuniformity} show the evolution of recall from two models presented in this paper, both of which are designed to be more robust to noise than the baseline.

First, Fig \ref{fig:nonstationarity-vs-nonuniformity} (middle) shows that the PCEN model, comprising per-channel energy normalization, is not only useful at dawn, but also earlier in the night: at certain sensor locations, the recall rate is above 70\% from 10 p.m. onwards, as opposed to 2 a.m. for the CNN baseline.
This qualitative finding confirms that replacing the logarithmic compression of the mel-frequency spectrogram (logmelspec) by per-channel energy normalization (PCEN) may turn out to be greatly beneficial to the practical usefulness of deep machine listening models for sound event detection.
Indeed, not only does PCEN significantly improve the tradeoff between precision and recall over the global test set (as was demonstrated in Fig \ref{fig:precision-recall}), but it also considerably reduces the probability of missed detection within time slots in which sound events are very rare, such as dusk in the case of avian flight calls.
It is striking to note that our end-to-end learning system, once endowed with a PCEN acoustic frontend, manages to perform almost as well on these time slots, despite the fact that, having fewer events, they contribute marginally to the global precision-recall curve of Fig \ref{fig:precision-recall}.

However, it should be noted that the increase in robustness to nonstationary noise that is afforded by the introduction of PCEN is not accompanied by an increase in robustness to nonuniform noise, as one could have hoped.
Rather, as illustrated by the shaded areas surrounding the line plots, and which denote interquartile variations across sensors, the GDA-PCEN model suffers from large variations in recall between sensor locations at any given time of the full night recording.
For example, near 2 a.m., the median recall for warblers and sparrows (frequency subband above \SI{5}{\kilo\hertz}) is about 60\%, but as low as 30\% for one of the sensors.
From the standpoint of the practitioner in the life sciences, the fact that such variations are both large and difficult to anticipate indicates that the use of PCEN alone is insufficient to offer any guarantees of reliability in the realm of automated bioacoustic event detection.

Secondly, Fig \ref{fig:nonstationarity-vs-nonuniformity} (bottom) shows the evolution of recall of the GDA-PCEN-AT model, also known as BirdVoxDetect, over the course of the BirdVox-full-night dataset.
It appears that this model, which combines a PCEN-based convolutional neural network and an auxiliary branch learning an adaptive threshold (AT), exhibits narrower interquartile differences between sensor locations than any of its counterparts.
In other words, even though context adaptation leaves the amount of robustness to nonstationarity in background noise essentially unchanged, it noticeably improves the robustness to nonuniformity in background noise of the sound event detection system at hand.
This observation suggests that the deployment of distributed machine listening software for flight call monitoring in a bioacoustic sensor network of autonomous recording units requires the resort to deep, data-driven methods for context adaptation, in addition to a shallow procedure of adaptive gain control in the time-frequency domain.

\section*{Conclusion}

Spatial and temporal variability in background noise and the inability to generalize automatic detectors in such conditions are major obstacles to the large-scale deployment of bioacoustic sensor networks.
In this article, we have developed, benchmarked, and combined several machine listening techniques to improve the generalizability of SED models across heterogeneous acoustic environments.

Our main finding is that, although both per-channel energy normalization (PCEN) and context adaptation (CA) improve the generalizability of deep learning models for sound event detection, these two methods are not interchangeable, but instead complementary:
whereas PCEN is best suited for mitigating the temporal variations of background noise in a single sensor, CA is best suited for mitigating spatial variations in background noise across sensor locations, whether the acoustic environment surrounding each sensor varies through time or not.
Indeed, PCEN relies on the assumption that background noise is stationary at a short time scale ($T_\mathrm{PCEN}=\SI{60}{\milli\second}$), of the order of the duration of the acoustic events of interest; whereas CA computes auxiliary features at a longer temporal scale ($T_\mathrm{CA}=30$ m).
Consequently, PCEN compensates intermittent changes in the loudness of background sources, such as a passing vehicle or the stridulation of an insect; however, it assumes statistical independence between background and foreground, and is thus inadequate to model how different habitats might trigger different vocalization behaviors in the species of interest. For its part, the CA-CNN draws on the variety of sensors in the training set to learn a joint model of both background and foreground; however, this joint model needs to be regularized by integrating long-term context into auxiliary features of relatively low dimensionality, which are, by design, invariant to rapid changes in environmental noise.

After a comprehensive benchmark of architectural variations between convolutional neural networks, we obtain statistically significant evidence to suggest that a combination of PCEN, adaptive threshold, and artificial data augmentation (pitch shifts and time stretchings) provides a consistent and interpretable improvement over the logmelspec-CNN baseline.
Reductions in miss rates with respect to the state of the art range between $10\%$ and $50\%$ depending on the location of the sensor, and bring the area under the precision-recall curve (AUPRC) of the BirdVox-full-night benchmark \cite{lostanlen2017icassp} from $61\%$ to $76\%$.
In addition, the recall of our selected model for sound event detection, named BirdVoxDetect, remains relatively high even in recording conditions where less training data is available, \eg{} at dusk or in sensors with peculiar characteristics in background noise.

Alongside this article, we release BirdVoxDetect as a pre-trained model on BirdVox-full-night.
We encourage bioacoustics researchers to download it and run it on their own recordings of nocturnal flight calls in the wild, especially if these recordings also contain high levels of background noise and/or spurious sound events.
BirdVoxDetect can detect the nocturnal flight calls of warblers, thrushes, and sparrows, with a high level of generality in terms of target species as well as sensor locations.
Indeed, as demonstrated by our benchmark, the procedures of PCEN and unsupervised context adaptation allow BirdVoxDetect to be deployed in a broad variety of recording conditions, exceeding those that are present in the BirdVox-full-night dataset.
The source code of BirdVoxDetect is freely available at the following URL address: \url{https://github.com/BirdVox/birdvoxdetect}.

Deriving computer-generated estimates of migratory activity at ranges of spatiotemporal scales from a decentralized network of low-cost bioacoustic sensors is a promising avenue for new insights in avian ecology and the conservation of biodiversity.
Future work will apply the BirdVoxDetect machine listening system to large-scale bioacoustic migration monitoring.

\section*{Acknowledgment}
We wish to thank Marc Delcroix, Holger Klinck, Peter Li, Richard F. Lyon, and Brian McFee for fruitful discussions.
We thank Thomas Grill and Jan Schl\"uter for sharing the source code of their ``bulbul'' system for bird detection in audio signals.
Lastly, we wish to thank the reviewers for their valuable comments and effort to improve the present manuscript.
This work is partially supported by NSF awards 1633259 and 1633206, the Leon Levy Foundation, and a Google faculty award.

\section*{Author contributions}

\begin{description}
\item[Conceptualization] VL, JS, AF, SK, JPB
\item[Data curation] AF, SK
\item[Formal analysis] VL
\item[Funding acquisition] AF, SK, JPB
\item[Investigation] VL, JS
\item[Methodology] VL, JS, JPB
\item[Project administration] AF, SK, JPB
\item[Resources] VL, JS
\item[Software] VL, JS
\item[Supervision] SK, JPB
\item[Visualization] VL, JS
\item[Writing --- original draft] VL, JS, AF
\item[Writing --- review \& editing] VL, AF
\end{description}

\nolinenumbers


%
%
%

\end{document}